\documentclass[12pt]{article}
\usepackage{amsmath}
\usepackage{amsfonts}
\usepackage{amssymb}
\usepackage{latexsym}
\usepackage{graphicx}
\begin{document}
\input epsf

\def\p{\partial}
\def\h{{1\over 2}}
\def\be{\begin{equation}}
\def\bea{\begin{eqnarray}}
\def\ee{\end{equation}}
\def\eea{\end{eqnarray}}
\def\d{\partial}
\def\la{\lambda}
\def\eps{\epsilon}
\def\bb{\bigskip}
\def\mm{\medskip}
\newcommand{\dm}{\begin{displaymath}}
\newcommand{\edm}{\end{displaymath}}
\renewcommand{\b}{\tilde{B}}
\newcommand{\gm}{\Gamma}
\newcommand{\ac}[2]{\ensuremath{\{ #1, #2 \}}}
\renewcommand{\ell}{l}
\newcommand{\z}{\ell}
\newcommand{\newsection}[1]{\section{#1} \setcounter{equation}{0}}
\def\bb{$\bullet$}

\def\q{\quad}

\def\bn{B_\circ}

\let\a=\alpha \let\b=\beta \let\g=\gamma \let\d=\delta \let\e=\epsilon
\let\c=\chi \let\th=\theta  \let\k=\kappa
\let\l=\lambda \let\m=\mu \let\n=\nu \let\x=\xi \let\r=\rho
\let\s=\sigma \let\t=\tau
\let\vp=\varphi \let\vep=\varepsilon
\let\w=\omega      \let\G=\Gamma \let\D=\Delta \let\Th=\Theta
                     \let\P=\Pi \let\S=\Sigma

\def\nn{\nonumber}
\let\bm=\bibitem

\let\pa=\partial

\begin{flushright}
OHSTPY-HEP-T-03-012\\
hep-th/yymmddd
\end{flushright}
\vspace{20mm}
\begin{center}
{\LARGE  Constructing `hair' for the three charge hole}
\\
\vspace{18mm}
{\bf   Samir D. Mathur\footnote{mathur@mps.ohio-state.edu},
Ashish Saxena\footnote{ashish@pacific.mps.ohio-state.edu} and Yogesh
Srivastava\footnote{yogesh@pacific.mps.ohio-state.edu}
\\}
\vspace{8mm}
Department of Physics,\\ The Ohio State University,\\ Columbus,
OH 43210, USA\\
\vspace{4mm}
\end{center}
\vspace{10mm}
\thispagestyle{empty}
\begin{abstract}

It has been found that the states of the 2-charge extremal D1-D5 system
are given by smooth geometries that
have no singularity and no horizon individually, but a `horizon' does
arise
after `coarse-graining'.  To see how this concept extends to the
3-charge extremal system, we construct a perturbation on the D1-D5
geometry that carries
one unit of momentum charge $P$. The perturbation is found to be
regular everywhere and normalizable,
so we conclude that at least this state of the 3-charge system behaves
like the 2-charge states. The solution is constructed by matching (to
several orders) solutions in the inner and outer regions of the
geometry.
We conjecture the general form of `hair' expected for the 3-charge
system, and the nature of the interior of black holes in general.

\end{abstract}
\newpage
\renewcommand{\theequation}{\arabic{section}.\arabic{equation}}
\newsection{Introduction}

The Bekenstein-Hawking entropy of a black hole is
\be
S={A\over 4G}
\label{one}
\ee
where $A$ is the area of the horizon. Statistical mechanics then
suggests
that the hole should have $e^{S}$ states. But where are these
states? In this paper we suggest  an answer to this question, and
support our conjecture by a calculation related to the 3-charge
extremal hole.

\subsection{Black hole `hair'}

String theory computations with extremal
and near extremal systems have shown that D-brane
states with the same charges and mass as the hole have precisely
$e^{S}$ states \cite{stromvafa, callanmalda}. If we increase the
coupling $g$ these states should give black holes \cite{susskind}. At
least for extremal holes supersymmetry tells us that we cannot gain or
lose
       any
states when we change $g$ \cite{vafa,sen}. We are thus forced
to address the question: How do the $e^S$ configurations
differ from each other in the gravity description?

Early attempts to find `hair' on black holes were based on looking for
small
perturbations in the metric and other fields while demanding smoothness
at the horizon. One found no such perturbations -- the energy in a
small deformation of the black hole solution would flow off to infinity
or fall into the singularity, and the hole would settle down to
its unique metric again.
But if we {\it had} found such hair at the horizon we would be faced
with an even more curious difficulty. We would have a set of
`microstates' as pictured in Fig.1(b), each looking like a black hole
but
differing slightly from other members of the ensemble.

\begin{figure}[ht]
\hspace{0.6in}
\includegraphics{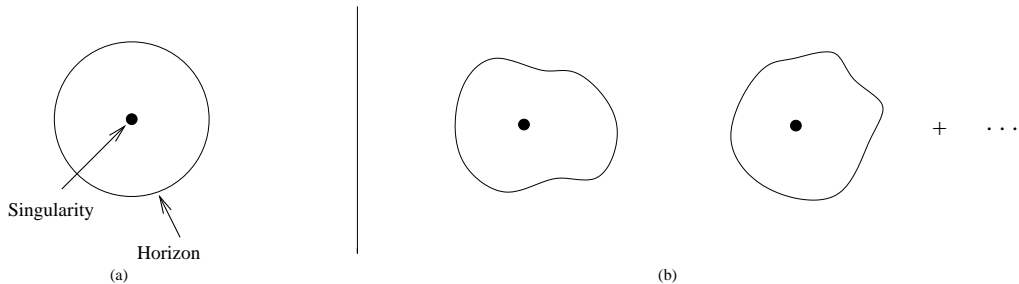}
\caption{ \small{(a) The usual picture of a black hole. ~(b) If the
microstates represented small deformations of (a) then each would
itself have
a horizon. }} \label{nfig1}
\end{figure}
But if each
microstate had a horizon as in the figure, then should'nt we assign an
entropy $\approx S $ to it? If we do, then we have $e^S$
configurations, with {\it each} configuration having an entropy 
$\approx S$.
This makes no sense -- we wanted the microstates to {\it explain} the
entropy, not have further entropy themselves. This implies that if we
do find the microstates in the gravity description, {\it then they
should
turn out to have no horizons themselves}.

We face exactly the same problem if we conjecture that the
configurations all look like Fig.1(a) but differ from each other near
the singularity;
each configuration would again have a horizon, and thus an entropy
$e^S$ of its own.

\begin{figure}[h]
\hspace{1.9in}
\includegraphics{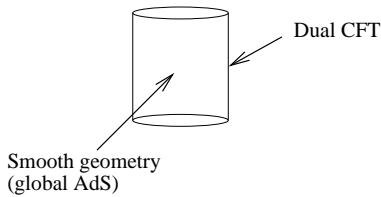}
\caption{ \small{The D1-D5 CFT is represented by a \emph{smooth}
geometry in the dual representation.}} \label{nfig2}
\end{figure}

The idea of $AdS/CFT$ duality \cite{maldacena} adds a further twist to
the problem. If
string states at weak coupling become black holes at larger coupling,
then one might think that the strings/branes are somehow sitting at the
center $r=0$ of the
black hole. The low energy dynamics of the branes is a CFT.  But  the 
standard description of AdS/CFT
duality says that the CFT is represented by a geometry that is {\it 
smooth} at
$r=0$   (Fig.\ref{nfig2}). In particular there are no sources or 
singularities
near $r=0$.

Putting all this together suggests the following requirements for black
hole `hair':
\medskip

(a)\quad There must be $e^{S}$ states of the hole.
\medskip

(b)\quad These individual states should have no horizon and no
singularity.
\medskip

(c)\quad `Coarse-graining' over these states should give the notion of
`entropy' for the black hole.
\bigskip

This appears to be rather an extreme change in our picture of the black
hole, particularly since (b) requires that the geometry of individual
states differ significantly from the standard black hole metric
everywhere in the interior of the hole, and not just within planck
distance of the singularity.

\begin{figure}[h]
\hspace{0.5in}
\includegraphics{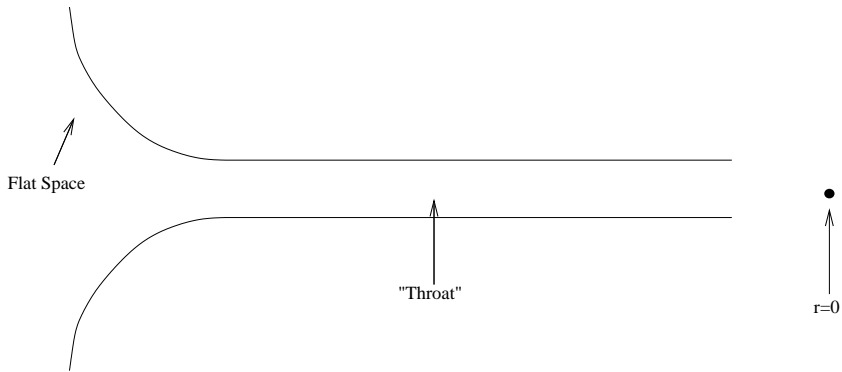}
\caption{ \small{The \emph{naive} geometry of the extremal D1-D5
system.  }} \label{nfig3}
\end{figure}

Remarkably though, just such a picture of individual states was found
for the 2-charge extremal D1-D5 system in \cite{lm4}\cite{lm5}. We take
$n_5$ D5 branes wrapped on $T^4\times S^1$ bound to $n_1$ D1 branes
wrapped on the
$S^1$. CFT considerations tell us that the entropy is
$S_{micro}=2\sqrt{2}\pi\sqrt{n_1n_5}$,  so the extremal ground state is
highly degenerate. In the gravity description we should see the same
number of configurations, except that in a classical computation this
degeneracy would show up as a continuous family of geometries rather
than discrete states. The {\it naive} metric that is usually written
down for
the D1-D5 state is pictured in Fig.\ref{nfig3} -- it goes to flat
space at
infinity,
and heads to a singularity at $r=0$.  But a detailed analysis shows the
following \cite{lm4, lm5}:

\medskip

(a$'$)\quad The actual classical geometry of the extremal D1-D5 system
is
found to be given by a family of states parametrized by a vector
function
$\vec F(v)$; upon quantization this family of geometries should yield
the $e^{2\sqrt{2}\pi\sqrt{n_1n_5}}$ states
expected from the entropy.

\medskip

\begin{figure}[h]
\hspace{0.5in}
\includegraphics{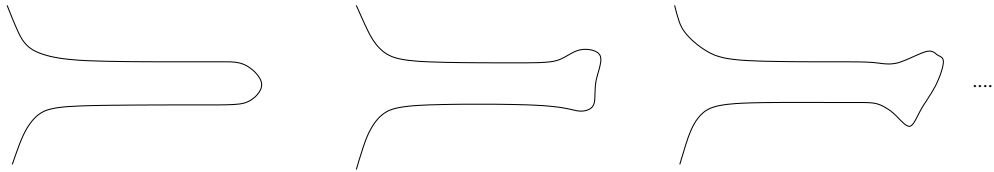}
\caption{ \small{\emph{Actual} geometries for different microstates of
the extremal D1-D5 system. }} \label{nfig4}
\end{figure}

(b$'$)\quad Individual members of this family of states have no horizon
and no singularity -- we picture this in Fig.\ref{nfig4}.

\medskip

\begin{figure}[t]
\hspace{1.0in}
\includegraphics{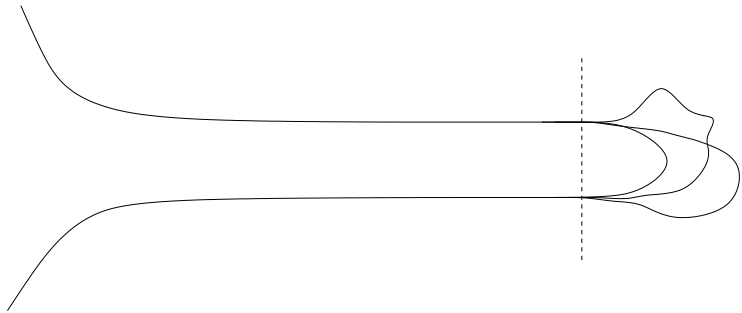}
\caption{ \small{Obtaining the `horizon' by `coarse-graining'.}}
\label{nfig5}
\end{figure}

(c$'$)\quad Suppose we define `coarse graining' for a family of
geometries in the following way. We draw a surface to separate the
region where the metrics are all essentially similar from the region
where they differ significantly from each other (indicated by the
dashed line in Fig.\ref{nfig5}). The area $A$ of this `horizon'
surface satisfies
\be
S\approx {A\over 4G}
\label{two}
\ee Note that the properties a$'$,b$'$,c$'$ address directly the
requirements
a,b,c.
\bigskip

\subsection{The three charge case}

The 2-charge D1-D5 extremal system has a `horizon' whose
radius is small compared to other length scales in the geometry, and
the entropy of this
system is determined from the geometry only upto a factor of order unity
(this is the reason for the $\approx$ sign in (\ref{two})). The
3-charge system which has D1, D5 and P charges
(P is momentum along $S^1$) has a  horizon radius that is of the same
order as other scales in the
geometry, and in the classical limit we get a Reissner-Nordstrom type
black hole. The
D-brane state entropy $S_{micro}$ exactly equals $S_{Bek}$
\cite{stromvafa}. We would
therefore like to find individual geometries that describe different
states of the 3-charge hole.
In line with what was said above, we expect a situation similar to that
in Figs.\ref{nfig3},\ref{nfig4} -- the {\it naive} D1-D5-P geometry has
a horizon at $r=0$, but
{\it actual} geometries end smoothly (without horizon or singularity)
before reaching $r=0$.

If this description of the 3-charge hole were  true then it would imply
a simple consequence: There should be  smooth perturbations of the
2-charge (D1-D5) system which add
a small amount of the third (momentum) charge. Thus we should find
small perturbations $\Psi$ around
the 2-charge geometries with the following properties

\medskip

(i)\quad The perturbation has momentum $p$ along the $S^1$, which
implies
\be
\Psi\sim e^{i{p\over R} y}, ~~~p\in \mathbb{Z}
\ee
where $y$ is the coordinate along  $S^1$ and $R$ is the radius of this 
$S^1$.

\medskip

(ii)\quad The perturbation takes the extremal 2-charge system to an
extremal 3-charge so the energy of the perturbation should equal the
momentum charge of the perturbation. This implies a $t$ dependence
\be
\Psi\sim e^{-i\omega t}, ~~\omega={p\over R}
\ee
\medskip

(iii)\quad The perturbation must generate no singularity and no
horizon, so it must be regular everywhere, and vanishing at
$r\rightarrow\infty$ so as to be normalizable.
\bigskip

\begin{figure}[ht]
\hspace{1.4in}
\includegraphics{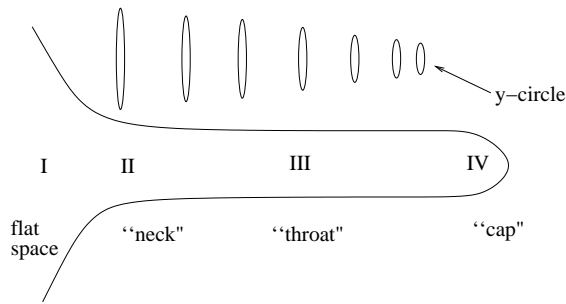}
\caption{ \small{Different regimes of the starting $2$-charge D1-D5
geometry. }} \label{nfig6}
\end{figure}

We start with a particular state of the 2-charge extremal system. We
have a  bound state of D1 and D5 branes,
wrapped on a $T^4$ with volume $(2\pi)^4V_4$ and an $S^1$ of radius 
$R$, sitting in asymptotically
flat $4+1$ transverse spacetime. This
system is in the Ramond (R) sector, which has many ground states. We
pick the particular one (we call it $|0\rangle_R$) which if spectral
flowed to the NS sector yields the NS vacuum $|0\rangle_{NS}$. The
geometry for this 2-charge state is pictured in
Fig.\ref{nfig6}. The radius of the $S^3$ in the region III is 
$(Q_1Q_5)^{1\over
4}$. The parameter
\be
\epsilon\equiv {(Q_1Q_5)^{1\over 4}\over R}
\ee
characterizes, roughly speaking,  the ratio ${{\rm diameter}\over {\rm
length}}$ for the `throat' region III.
\medskip

In the NS sector we can act with a chiral primary operator on
$|0\rangle_{NS}$.
Let the resulting state  be called
$|\psi\rangle_{NS}$. The spectral flow of this state to the R sector
gives a state
$|\psi\rangle_R$; this will be an R ground state, and will have
$L_0=\bar L_0={c\over 24}$.
We will construct the perturbation that will describe the CFT state
\be
(J_{-1}^-)|\psi\rangle_R
\label{three}
\ee
This state has momentum charge $L_0-\bar L_0=1$.
We proceed in the following steps:

\bigskip
(A)\quad The regions III and IV are actually a part of global 
$AdS_3\times
S^3\times T^4$, and a coordinate change brings the
metric here to the standard form \cite{bal, mm}. The wavefunction
$\Psi_{inner}$ for the state (\ref{three}) in this region can be
obtained by rotating a chiral primary perturbation in global
$AdS_3\times S^3$.

\medskip

(B)\quad We construct the appropriate wavefunction $\Psi_{outer}$ in
the regions I, II, III by solving the supergravity equations
in this part of the geometry. We choose a solution that decays at
infinity.
\medskip

(C)\quad We find that at leading order  $\epsilon^0$ the solutions
$\Psi_{inner}$, $\Psi_{outer}$ agree in the overlap region III.
\medskip

(D)\quad We extend the computation to order $\epsilon, \epsilon^2,
\epsilon^3$ and
continue to find agreement in the overlap region;
this agreement appears to be highly nontrivial, and we take it as
evidence for the existence of the solution satisfying
(i), (ii), (iii) above.

\medskip

After this computation we conclude with some conjectures about the form
of `hair' for generic states of the 3-charge hole,
and a discussion of the physics underlying the new picture of the black
hole interior that emerges from this structure of microstates.

\newsection{The 2-charge system: review}

In this section we review the results obtained earlier for the 2-charge
D1-D5 system and
describe the particular D1-D5 background to which we will add the
perturbation carrying momentum charge P.

\subsection{Generating the `correct' D1-D5 geometries}

Consider IIB string theory compactified on $T^4\times S^1$. The D1-D5
system can be mapped by a set of $S,T$ dualities to the FP system
\begin{eqnarray}n_5 ~D5 ~{\rm branes~along} ~T^4\times
S^1 &\rightarrow& n_5~ {\rm units ~of ~fundamental ~string
~winding~along}~S^1~ (F)\nonumber\\
n_1~ D1~ {\rm branes~along}~S^1 &\rightarrow& n_1~ {\rm units
~of~ momentum~along}~S^1~ (P)\nonumber
\end{eqnarray}
The {\it naive} metric of the FP bound state in string frame is
\be
ds^2=-(1+{Q\over r^2})^{-1}(dudv+{Q'\over r^2}dv^2)+dx_idx_i+dz_adz_a
\label{naivefp}
\ee
where $x_i, i=1\dots 4$ are the noncompact directions, $z_a, a=1\dots
4$ are the $T^4$ coordinates, and we have smeared all functions on
$T^4$. We will also use the definitions
\be
u=t+y, ~~~v=t-y
\ee

But in fact the bound state of the F and P charges corresponds to a
fundamental string `multiwound' $n_5$ times
around $S^1$, with all the momentum P being carried on this string as
traveling waves. Since the F string has no longitudinal
vibrations, these waves necessarily cause the strands of the multiwound
string to bend away and separate from each other in the transverse
directions. The possible configurations are parametrized by  the
transverse displacement $\vec F(v)$; we let this vibration be only in
the noncompact directions $x_1, x_2, x_3, x_4$. The resulting solution
can be constructed using the
techniques of \cite{tsey,dabholkar,callan}, and we find for the metric
in string frame
\cite{lm3}\footnote{We
can extend the construction to get additional states by letting the
string vibrate along the $T^4$ directions; these states were
constructed in \cite{lmm}.}${}^,$\footnote{The angular momentum bounds
of
\cite{lm3} and metrics found
in \cite{bal, mm, lm3} were reproduced in the duality related F-D0
system through `supertubes' \cite{supertubes}.
While supertubes help us to understand some features of the physics we
still find that to construct metrics
of general bound states of 2-charges and to identify the metrics with
their CFT dual states the best way
is to start with the FP system.}
\bea
ds^2&=&H(-dudv+K dv^2+2A_idx_i dv)+dx_idx_i+dz_adz_a\nonumber\\
B_{vu}&=&-G_{vu}=\h H, ~~B_{vi}=-G_{vi}=-H A_i, ~~e^{-2\Phi}=H^{-1}
\eea
where
\be
H^{-1}=1+{Q\over L}\int_0^L {dv\over |\vec x-\vec F(v)|^2}, ~~K={Q\over
L}\int_0^L {dv (\dot
F(v))^2\over |\vec x-\vec F(v)|^2},
~~A_i=-{Q\over L}\int_0^L {dv\dot F_i(v)\over |\vec x-\vec F(v)|^2}\\
\label{functions}
\ee
($L=2\pi n_1 R$, the total length of the F string.\footnote{Parameters
like $Q,R$ are not the same for the FP and D1-D5 systems -- they are
related by duality transforms. Here we have not used different symbols
for the two systems
to avoid cumbersome notation and the context should clarify what the
parameters  mean. For full details on the relations between parameters
see \cite{lm3,lm4}).})

Undoing the S,T dualities we find the solutions describing the family
of Ramond ground states of the D1-D5 system \cite{lm4}
\be
ds^2=\sqrt{H\over 1+K}[-(dt-A_i dx^i)^2+(dy+B_i dx^i)^2]+\sqrt{1+K\over
H}dx_idx_i+\sqrt{H(1+K)}dz_adz_a
\label{six}
\ee
\bea
e^{2\Phi}&=&H(1+K), ~~~C^{(2)}_{ti}={B_i\over 1+K},
~~~C^{(2)}_{ty}=-{K\over 1+K}\nonumber\\
C^{(2)}_{iy}&=&-{A_i\over 1+K},
~~~~C^{(2)}_{ij}=C_{ij}+{A_iB_j-A_jB_i\over 1+K}
\label{twenty}
\eea
where $B_i, C_{ij}$ are given by
\be
dB=-*_4dA, ~~~dC=-*_4dH^{-1}
\ee
and $*_4$ is the duality operation in the 4-d transverse  space
$x_1\dots
x_4$ using the flat metric $dx_idx_i$. The functions $H^{-1}, K, A_i$
are
the same as the functions in (\ref{functions})

It may appear the the solution (\ref{six}) will be singular at the
points $\vec x=\vec F(v)$, but it was found in \cite{lm4} that
this singularity reflects all incoming waves in a simple way. The
explanation for this fact was pointed out in a nice calculation in
\cite{lmm} where it was shown that the singularity  (for generic $\vec
F(v)$) is a {\it coordinate} singularity; it is the same coordinate
singularity as the one encountered at the origin of a Kaluza-Klein
monopole \cite{grossperry}.

The family of geometries (\ref{six}) thus have the form pictured in
Fig.\ref{nfig4}. These geometries are to be contrasted with  the
`naive'
geometry for the D1-D5 system
\be
ds^2_{naive}={1\over \sqrt{(1+{Q_1\over r^2})(1+{Q_5\over
r^2})}}[-dt^2+dy^2]+\sqrt{(1+{Q_1\over r^2})(1+{Q_5\over
r^2})}dx_idx_i+\sqrt{{1+{Q_1\over r^2}\over 1+{Q_5\over r^2}}}dz_adz_a
\label{naive}
\ee
The actual geometries (\ref{six}) approximate this naive geometry
everywhere except near the `cap'.

It is important to note that
we can perform dynamical experiments with these different geometries
that distinguish them from each other. In \cite{lm4} the travel time
$\Delta t_{sugra}$ was computed for a waveform to travel down and back
up
the `throat' for a 1-parameter family of such geometries. Different
geometries in the family had different lengths for the `throat' and
thus different $\Delta t_{sugra}$. For each geometry we found
\be
\Delta t_{sugra}=\Delta t_{CFT}
\label{equal}
\ee
where   $\Delta t_{CFT}$ is the time taken for the corresponding
excitation to travel once around the `effective string' in the  CFT
state dual to the given geometry. Furthermore, the backreaction of the
wave on the geometry was computed and shown to be small so that the
gravity computation made sense.

In \cite{lm5} a  `horizon' surface was constructed to separate the
region where the geometries agreed with each other from the region
where they differed, and it was observed that the entropy of
microstates agreed with the Bekenstein entropy that one would associate
to this surface\footnote{In \cite{sen} the naive geometry for FP was
considered, and
it was argued that since the curvature became order string scale below
some $r=r_0$, a
`stretched horizon' should be placed at $r_0$. The area $A$ of this
stretched horizon
also satisfied ${A\over 4G}\sim S_{micro}$.  It is unclear, however,
how this criterion
for a `horizon' can be used for the duality related D1-D5 system, where
the geometry for small
$r$ is locally $AdS_3\times S^3$ and the curvature is {\it constant}
(and small). We, on the other
hand have observed that geometries for different microstates {\it
depart} from each other for $r\le r_0$ and placed the horizon
at this location; this gives the same horizon location for all  systems
related by duality.}
\be
S_{micro}\sim {A\over 4G}
\ee
Such an  agreement was also found  for the 1-parameter family of  
`rotating
D1-D5 systems' where the states in the system were constrained to have
an angular momentum $J$. The horizon surfaces in these  cases had
the shape of a `doughnut'.

\subsection{The geometry for $|0\rangle_R$}

The geometry dual to the R sector state $|0\rangle_R$ (which
results from the spectral flow of the
NS vaccum $|0\rangle_{NS}$) is found by starting with the FP profile
\be
f_1(v)=a\cos({v\over n_5 R}), ~~f_2(v)=a\sin({v\over n_{5 }R}),
~~f_3(v)=0, ~~f_4(v)=0
\ee
and constructing the corresponding D1-D5 solution. The geometry for
this case had arisen earlier in different studies
in \cite{cy, bal, mm}. For simplicity we set
\be
Q_1=Q_5\equiv Q
\ee
which gives the D1-D5 solution
\bea
ds^2&=&-{1\over h}(dt^2-dy^2)+hf(d\theta^2+{dr^2\over
r^2+a^2})-{2aQ\over hf}(\cos^2
\theta dy d\psi+\sin^2\theta dt d\phi)\nonumber\\
&+&h[(r^2+{a^2Q^2\cos^2\theta\over h^2f^2})\cos^2\theta
d\psi^2+(r^2+a^2-{a^2Q^2\sin^2\theta\over h^2f^2})\sin^2\theta
d\phi^2]+dz_adz_a\nonumber\\
\label{el}
\eea
where
\be
a={Q\over R}, ~~~f=r^2+a^2\cos^2\theta, ~~~h=1+{Q\over f}
\ee
The dilaton and RR field are
\bea
e^{2\Phi}&=&1, ~~~~C^{(2)}_{ty}=-{Q\over Q+f},
~~~~C^{(2)}_{t\psi}=-{Qa\cos^2\theta\over Q+f}\nonumber\\
C^{(2)}_{y\phi}&=&-{Qa\sin^2\theta\over
Q+f},~~~~C^{(2)}_{\phi\psi}=Q\cos^2\theta+{Qa^2\sin^2\theta\cos^2\theta
\over Q+f}\label{selfdualfield}
\eea
To construct the 3-charge solution below we will assume that
\be
\epsilon\equiv {a\over \sqrt{Q}}={\sqrt{Q}\over R}<<1
\ee
which can be achieved by taking the compactification radius $R$ to be
large for fixed values of $\alpha', g, n_1, n_5, V_4$.
  In what follows we will
ignore the $T^4$ and write 6-d metrics only. Since
the dilaton $\Phi$ and $T^4$ volume are constant in the above solution
the 6-d Einstein metric is the same as the 6-d string metric.

\subsubsection{The `inner' region}

For
\be
r<<\sqrt{Q}
\label{eightt}
\ee
the geometry (\ref{el}) becomes
\bea
ds^2&=&-{(r^2+a^2\cos^2\theta)\over
Q}(dt^2-dy^2)+Q(d\theta^2+{dr^2\over r^2+a^2})\nonumber\\
&&~~~
-2a(\cos^2\theta dyd\psi+\sin^2\theta dtd\phi)+Q(\cos^2\theta
d\psi^2+\sin^2\theta d\phi^2)
\label{innerr}
\eea
The change of coordinates
\be
\psi_{NS}=\psi-{a\over Q}y, ~~~\phi_{NS}=\phi-{a\over Q}t
\label{spectral}
\ee
brings (\ref{innerr}) to the form $AdS_3\times S^3$
\be
ds^2=-{(r^2+a^2)\over Q}dt^2+{r^2\over Q}dy^2+Q{dr^2\over
r^2+a^2}+Q(d\theta^2+\cos^2\theta d\psi_{NS}^2+\sin^2\theta
d\phi_{NS}^2)
\label{innerns}
\ee

We will call the region (\ref{eightt}) the {\it inner region} of the
complete geometry (\ref{el}).

\subsubsection{The `outer' region}

The region
\be
a<<r<\infty
\label{sevent}
\ee
is flat space ($r\rightarrow\infty$) going over to the `Poincare patch'
(with $y\rightarrow y+2\pi R$ identification)
\be
ds^2=-{r^2\over Q+r^2}(dt^2-dy^2)+(Q+r^2){dr^2\over
r^2}+(Q+r^2)[d\theta^2+\cos^2\theta d\psi^2+\sin^2\theta
d\phi^2]
\label{outer}
\ee

We will call the
region (\ref{sevent}) the {\it outer region} of the geometry (\ref{el}).
The inner and outer regions have a domain of overlap
\be
a<<r<<\sqrt{Q}
\label{nint}
\ee

\subsubsection{The spectral flow map}

The coordinate transformation (\ref{spectral}) taking (\ref{innerr}) to
(\ref{innerns}) gives {\it spectral flow} \cite{bal, mm}. The fermions
of
the supergravity theory are periodic around the $S^1$ parametrized by
the coordinate $y$ in (\ref{innerr}), but the transformation
(\ref{spectral})
causes the $S^3$ to rotate once as we go around this $S^1$, and the
spin of the fermions under the rotation group of this $S^3$ makes them
antiperiodic around $y$ in the metric (\ref{innerns}). Thus the metric
(\ref{innerr}) gives the dual field theory in the R sector while the
metric
(\ref{innerns}) describes the CFT in the NS sector.

\newsection{The  perturbation carrying momentum}

\subsection{The equations}

The fields of IIB supergravity in 10-d give rise to a large number of
fields after reduction to 6-d. At the same time we
get an enhancement of the symmetry group, as various different fields
combine into larger representations of the  6-d
theory.\footnote{In the actual reduction of IIB from 10-d to 6-d we
also get additional fields like $A_\mu\equiv h_{a\mu}$, where $a=1\dots
4$ is a $T^4$ direction. We do not study these additional fields here.
} In \cite{sezgin} general 4b supergravities in 6-d were studied around
$AdS_3\times S^3$; their perturbation equations however apply to the
more general
background that we will use. These supergravities
have the graviton $g_{MN}$, self-dual 2-form fields $C^i_{MN}, i=1\dots
5$, anti-self-dual 2-forms $ B^r_{MN}, r=1\dots n$ and scalars
$\phi^{ir} $.

Suppose we have a solution to the field equations with a nontrivial
value
for the metric and one of the self-dual fields
\be
g_{MN}=\bar g_{MN}, ~~~C^1_{MN}=\bar C^1_{MN}\equiv C_{MN}
\label{tone}
\ee
The choice $Q_1=Q_5=Q$ has made the  field $C^{(2)}$ in
(\ref{selfdualfield}) self-dual, and gives us such a background.
(This choice simplifies the computations, but we expect that the
perturbation we are constructing will exist for
general $Q_1, Q_5$ as well.)

Linear perturbations around the background (\ref{tone}) separate into
different  sets. The  anti-self-dual
field $B^r_{MN}$ mixes only with the scalar $\phi^{1r}$. We set $r=1$
using the $SO(n)$ symmetry of the theory and write
\be
B^1_{MN}\equiv B_{MN}, ~~F_{MNP}=\p_MB_{NP}+\p_NB_{PM}+\p_PB_{MN},
~~~\phi^{11}\equiv w
\ee

The field equations are\footnote{Our 2-form fields are twice the 2-form 
fields in \cite{sezgin}. Our normalizations agree with those 
conventionally used for the
10-D supergravity fields where the action is $-{1\over 12} \int F^2$.} 
(we write $\bar H_{MNP}=\p_M\bar C_{NP}+\p_N\bar
C_{PM}+\p_P\bar C_{MN}$)
\be
F_{ABC}+{1\over 3!}\epsilon_{ABCDEF}F^{DEF}+w\bar H_{ABC}=0 \label{eom1}
\ee
\be
w_{;A}{}^{;A}-{1\over 3}\bar H^{ABC}F_{ABC}=0\label{eom2}
\ee

\subsection{The $(B,w)$ perturbation at leading order
($O(\epsilon^0)$)}

In this subsection we construct the desired perturbation to leading
order
in the inner and outer regions
and observe their agreement at this order of approximation.

\subsubsection{Inner region: The chiral primary $|\psi\rangle_{NS}$}

Consider the equations (\ref{eom1}),(\ref{eom2}) in the inner region.
In the coordinates  (\ref{innerns}) this region
is seen to
be just `global' $AdS_3\times S^3$. We use $a,b\dots$ to denote indices
on $S^3$ and $\mu, \nu\dots$ to denote indices
on $AdS_3$.  We find the following solution  for these
equations in global $AdS_3\times S^3$
\be
w={e^{-2i{a\over Q}l t}\over Q (r^2+a^2)^l}{\hat Y}_{NS}^{(l)}
\label{ponepre}
\ee
\be
B_{ab}=B\epsilon_{abc}\partial^c{\hat Y}^{(l)}_{NS},
~~~B_{\mu\nu}={1\over
\sqrt{Q}}\epsilon_{\mu\nu\lambda}\p^\lambda B ~{\hat Y}^{(l)}_{NS}
\label{pone}
\ee
where
\be
{\hat Y}^{(l)}_{NS}=(Y^{(l,l)}_{(l,l)})_{NS}= \sqrt{ \frac{2l+1}{2
} }\frac{
e^{-2il\phi_{NS}}}{\pi} \sin^{2l}\theta, ~~~B={1\over 4l}{e^{-2i {a\over Q}l t }\over
(r^2+a^2)^l}
\label{ponepost}
\ee
In (\ref{pone}) the tensors $\epsilon_{abc}, g^{ab}$
etc are defined using the metric
on an $S^3$ with {\it unit} radius. This choice simplifies the
presentation of spherical harmonics but results in the factor
$({\rm radius ~of~}S^3)^{-1}={1\over \sqrt{Q}}$ in the definition of
$B_{\mu\nu}$.
The tensors $\epsilon_{\mu\nu\lambda}, g^{\mu\nu}$ etc. are defined
using the
$t,y,r$ part of the metric (\ref{innerns}).\footnote{The spherical
harmonics are representations of $so(4)\approx su(2)\times su(2)$;
the upper labels in $Y^{(l,l)}_{(l,l)}$ give the $j$ values in each
$su(2)$, and the lower indices give the $j_3$ values.
Thus $l=0, {1\over 2}, 1, \dots$.
The subscript $NS$  on $Y$ indicates that the arguments are the sphere
coordinates in the NS sector, $(\theta, \psi_{NS}, \phi_{NS})$. When we
write no such subscript it is to be assumed that the arguments of the
spherical harmonic are the R sector coordinates $(\theta, \psi, \phi)$.
More details about spherical harmonics are given in Appendix A.}

This solution represents a chiral primary of the dual CFT
\cite{exclusion}. To see this
note the AdS/CFT relations
giving charges and dimensions of bulk excitations
\be
J^{NS}_z={i\over 2}[\p_{\psi_{NS}}+\p_{\phi_{NS}}], ~~~{\bar
J}_z^{NS}={i\over 2}[-\p_{\psi_{NS}}+\p_{\phi_{NS}}]
\ee
\be
L_0^{NS}=i{Q\over a}\p_u, ~~~{\bar L}_0^{NS}=i{Q\over a}\p_v
\ee
The solution (\ref{ponepre})-(\ref{ponepost}) thus has
\be
j^{NS}=l, ~~h^{NS}=l, ~~~{\bar j}^{NS}=l,  ~~{\bar h}^{NS}=l
\ee
which are the conditions for a chiral primary.

The coordinate transformation (\ref{spectral}) brings us to the R
sector. The
scalar in these coordinates is
\be
w={1\over Q (r^2+a^2)^l}{\hat Y}^{(l)}, ~~~~{\hat
Y}^{(l)}= \sqrt{\frac{2l+1}{2 }} \frac{e^{-2il\phi}}{\pi} 
\sin^{2l}\theta
\ee
so that it has no $t$ or $y$ dependence. The components of $B_{AB}$
similarly do not
have any $t,y$ dependence.

The dimensions in the R sector are given by (the partial derivatives 
this time are with respect to the R sector variables)
\be
L_0=i{Q\over a}\p_u, ~~~{\bar L}_0=i{Q\over a}\p_v
\ee
so that we get for our perturbation
\be
h=\bar h=0
\ee
which is expected, since a chiral primary of the NS sector maps under
spectral flow to a ground state of the R
sector.\footnote{The full spectral flow relations are
$h=h_{NS}-j_{NS}+{c\over 24}, ~j=j_{NS}-{c\over 12}$. Spectral flow of
the
background $|0\rangle_{NS}$ gives $h^0=h^0_{NS}-{c\over 24},
~j^0=j^0_{NS}-{c\over 12}$, so for
the
perturbation the spectral flow relations are just
$h=h_{NS}-j_{NS}, j=j_{NS}$.}

Let the CFT state dual to the perturbation
(\ref{ponepre})-(\ref{ponepost}) be called $|\psi\rangle_{NS}$, and
let $|\psi\rangle_R$ be its image under spectral flow to the Ramond
sector.

\subsubsection{Inner region: The state
$J^-_0|\psi\rangle_{NS}~\leftrightarrow~J^-_{-1}|\psi\rangle_R$}

      Consider again the inner region in the NS sector
coordinates (\ref{innerns}). We now wish to make the perturbation dual
to the NS sector state
\be
J^-_0|\psi\rangle_{NS}
\ee
Since the operator $J^-_0$ in the NS sector is represented by just a
simple rotation of the $S^3$, we can immediately write down the bulk
wavefunction dual to the above CFT state
\be
w={e^{-2i{a\over Q}l t}\over Q (r^2+a^2)^l}Y^{(l)}_{NS}
\label{solone}
\ee
\be
B_{ab}=B\epsilon_{abc}\p^c Y^{(l)}_{NS},
~~~B_{\mu\nu}={1\over
\sqrt{Q}}\epsilon_{\mu\nu\lambda}\partial^\lambda B~ Y^{(l)}_{NS}
\label{soltwo}
\ee
\be
Y^{(l)}_{NS}=(Y^{(l,l)}_{(l-1,l)})_{NS}= -\frac{\sqrt{l(2l+1)}}{\pi}
\sin^{2l-1}\theta \cos\theta
e^{i(-2l+1)\phi_{NS}+i\psi_{NS}},
~~~B={1\over 4l}{e^{-2i {a\over Q}l t }\over
(r^2+a^2)^l}
\label{solthree}
\ee
This perturbation has
\be
j_{NS}=l-1, ~~\bar j_{NS}=l, ~~~h_{NS}=l, ~~\bar h_{NS}=l
\label{nonmatch}
\ee
The spectral flow to the R sector coordinates should give
\be
h=h_{NS}-j_{NS}=1, ~~~\bar h=\bar h_{NS}-\bar j_{NS}=0
\ee
so that we have a state with nonzero $L_0-\bar L_0$, which means that
it is a state with momentum.
This can be seen explicitly by writing the solution
(\ref{solone})-(\ref{solthree})
in the R sector coordinates. Writing
\be
Y^{(l)}=
-\frac{\sqrt{l(2l+1)}}{\pi} e^{i(-2l+1)\phi+i\psi}\sin^{2l-
1}\theta\cos\theta, ~~~u=t+y
\label{choice}
\ee
we get
\bea
w&=&{1\over Q}{e^{-i{a\over Q}u}\over (r^2+a^2)^l}Y^{(l)}
\label{first1} \\
B_{\theta\psi} &=& \frac{1}{4 l}\frac{e^{-i\frac{a}{Q} u}}{(r^2+a^2)^l}
\cot\theta \partial_{\phi}Y^{(l)} \\
B_{\theta\phi} &=& -\frac{1}{4 l}\frac{e^{-i\frac{a}{Q}
u}}{(r^2+a^2)^l} \tan\theta \partial_{\psi}Y^{(l)} \\
B_{\psi\phi} &=& \frac{1}{4 l}\frac{e^{-i\frac{a}{Q} u}}{(r^2+a^2)^l}
\sin\theta\cos\theta \partial_{\theta}Y^{(l)} \\
B_{t \theta} &=& -\frac{a}{4 l}\frac{e^{-i\frac{a}{Q} u}}{Q(r^2+a^2)^l}
\tan\theta \partial_{\psi}Y^{(l)} \\
B_{t \psi} &=& \frac{a}{4 l}\frac{e^{-i\frac{a}{Q} u}}{Q(r^2+a^2)^l}
\sin\theta\cos\theta \partial_{\theta}Y^{(l)}
\eea
\bea
B_{y \theta} &=& \frac{a}{4 l}\frac{e^{-i\frac{a}{Q} u}}{Q(r^2+a^2)^l}
\cot\theta \partial_{\phi}Y^{(l)} \\
B_{y \phi} &=& -\frac{a}{4 l}\frac{e^{-i\frac{a}{Q} u}}{Q(r^2+a^2)^l}
\sin\theta\cos\theta \partial_{\theta}Y^{(l)} \\
B_{ty} & = & - \frac{1 }{2 Q^{2} } \frac{r^{2} e^{-i \frac{a}{Q} u}
}{(a^2 + r^2 )^{l} } Y^{(l)} \\
B_{yr} &=& \frac{i }{2Q} \frac{r e^{-i \frac{a}{Q} u} }{(r^2+a^2)^{l+1}
} Y^{(l)}\label{last1}
\eea

We see that all fields behave as $\sim e^{-i\omega t + i\lambda y}$
with $\omega=|\lambda|$,
so we have a BPS perturbation adding a third charge (momentum P$=-1$)
to the
2-charge
D1-D5 background.

\subsubsection{Outer region: Continuing the perturbation
$J^-_{-1}|\psi\rangle_R$}

We now wish to ask if this solution in the inner region continues out
to asymptotic infinity, falling off in a way
that makes it a normalizable perturbation. To do this we solve the
perturbation equations (\ref{eom1}),(\ref{eom2}) in the outer region
(\ref{outer}).
Requiring decay at infinity, we find the solution
\be
w={e^{-i{a\over Q} u}\over (Q+r^2)r^{2l}}Y^{(l)}
\label{solrone}
\ee
\be
B_{ab}=B\epsilon_{abc}\partial^c Y^{(l)}, ~~~B_{\mu\nu}={1\over
\sqrt{Q+r^2}}\epsilon_{\mu\nu\lambda}\partial^\lambda B~ Y^{(l)},
~~~B={1\over
4l}{e^{-i{a\over Q} u}\over r^{2l}}
\label{solrtwo}
\ee
where we have chosen the same spherical harmonic $Y^{(l)}$ that appears
in (\ref{choice}). Again $\epsilon_{abc}, g^{ab}$ etc.
refer to the metric of a {\it unit} $S^3$ (this gives  the factor $({\rm
radius ~of~}S^3)^{-1}={1\over \sqrt{Q+r^2}}$ in $B_{\mu\nu}$), while
$\epsilon_{\mu\nu\lambda}, g^{\mu\nu}$ etc. refer to the $t,y,r$ part
of the metric (\ref{outer}). Writing explicit components, the above 
solution
becomes
\bea
w&=&{e^{-i{a\over Q} u}\over (Q+r^2)r^{2l}}Y^{(l)}\label{first1p}\\
B_{\theta\psi} &=& \frac{1}{4 l}\frac{e^{-i\frac{a}{Q} u}}{r^{2l}}
\cot\theta \partial_{\phi}Y^{(l)} \\
B_{\theta\phi} &=& -\frac{1}{4 l}\frac{e^{-i\frac{a}{Q} u}}{r^{2l}}
\tan\theta \partial_{\psi}Y^{(l)} \\
B_{\psi\phi} &=& \frac{1}{4 l}\frac{e^{-i\frac{a}{Q} u}}{r^{2l}}
\sin\theta\cos\theta \partial_{\theta}Y^{(l)} \eea
  \bea
B_{ty} & = & -\frac{1  }{2 (Q+r^2)^{2} } \frac{ e^{-i \frac{a}{Q} u}
}{r^{2l-2} } Y^{(l)} \\
B_{tr} &=&\frac{i a }{r^{2l+1} }
\frac{1}{4 l Q  }e^{-i \frac{a}{Q} u}Y^{(\z)}\\
B_{yr} &=&  \frac{i  a}{r^{2l+1}}
\frac{1}{4\ell Q  } e^{-i\frac{a}{Q}u}Y^{(\z)}
\eea

\subsubsection{Matching at leading order}

We wish to see if the solutions in the inner and outer regions agree in
the domain of overlap
$a<<r<<Q$. In this region we have
\be
{a\over \sqrt{Q}}~<<~\{{a\over r}, {r\over \sqrt{Q}}\} << 1
\ee
We can match the solutions around any $r$ in the range $a<<r<<Q$. To
help us organize our perturbation expansion we choose this matching
region to be around the  geometric mean of $a,Q$, so that
\be
{a\over r} ~\sim ~ {r\over \sqrt{Q}}~\sim ~\epsilon^{1\over 2}
\ee

In this region the scalar $w$ in the inner region  (given by
(\ref{first1}))
and in the outer region (given by (\ref{first1p})) both reduce to the
same
function
\be
w={e^{-i{a\over Q} u}\over Qr^{2l}}Y^{(l)}+\dots
\ee
so that we get the desired agreement at leading order. We can similarly
compare $B_{MN}$, but note that since $B_{MN}$ is a tensor the 
components of $B_{MN}$
depend on the coordinate frame.  To see the order of a given component
$B_{MN}$ we should construct the field strength $F=dB$ from this
component and then look at the values of $F$ in an orthonormal frame.
For example
\be
B_{ty}\rightarrow  F_{\hat t\hat y\hat r}\equiv F_{tyr}(g^{tt})^{1\over
2}(g^{yy})^{1\over 2})(g^{rr})^{1\over 2}\sim {1\over Q^{3\over 2}
r^{2l}}
\ee
Note that $\bar H_{\hat t\hat y\hat r}\sim {1\over \sqrt{Q}}$, so that 
the
$F_{\hat t\hat y\hat r}$ in the above equation is of the same order as
$w\bar H_{\hat t\hat y\hat r}$, and thus $B_{ty}$  is a term which we
will match at leading order.

We then find that the components surviving at leading order  reduce to
the following forms for both the inner and outer solutions
\bea
B_{\theta\psi} &=& \frac{1}{4 l}\frac{e^{-i\frac{a}{Q} u}}{r^{2l}}
\cot\theta \partial_{\phi}Y^{(l)} \\
B_{\theta\phi} &=& -\frac{1}{4 l}\frac{e^{-i\frac{a}{Q}
u}}{r^{2l}} \tan\theta \partial_{\psi}Y^{(l)} \\
B_{\psi\phi} &=& \frac{1}{4 l}\frac{e^{-i\frac{a}{Q} u}}{r^{2l}}
\sin\theta\cos\theta \partial_{\theta}Y^{(l)}
\eea
\bea
B_{ty} & = & - \frac{1 }{2 Q^{2} } \frac{r^{2} e^{-i \frac{a}{Q} u}
}{r^{2l}} Y^{(l)}
\eea
Other components like $B_{y\phi}$ which do not agree
are seen to be higher order terms. We will find agreement for these
after we correct the inner and outer region computations by higher
order terms.

\subsection{Nontriviality of the matching}

Before proceeding to study the solutions and matching at higher orders
in $\epsilon$, we observe that the above  match  at
leading order is itself nontrivial. The dimensional reduction from 10-d
to 6-d also gives some massless scalars in 6-d
\be
\square s=0
\label{scalareq}
\ee
We show that for such a scalar we {\it cannot} get any solution that is
regular everywhere and decaying at infinity.
For the scalar $s$ we can find in the inner region $AdS_3\times S^3$ a
solution analogous to (\ref{solone}) \cite{lmpp}
\be
s={e^{-i(2l+2) {a\over Q} t}\over (r^2+a^2)^{l+1}}{ Y}^{(l)}_{NS}
\label{scalarsol}
\ee
where we have chosen the same spherical harmonic  as in (\ref{solone}).
Since the scalar generates not a chiral primary but a supersymmetry 
descendent, we
get instead of  (\ref{nonmatch})
\be
j_{NS}=l-1, ~~\bar j_{NS}=l, ~~~h_{NS}=l+1, ~~\bar h_{NS}=l+1
\ee
The solution (\ref{scalarsol}) falls off towards the boundary of
$AdS$, but in the complete geometry
(\ref{el}) it will not be normalizable at infinity. Using  R sector
coordinates
(which are natural at $r\rightarrow\infty$) we find that the $t,y$
dependence  is
$e^{-i\omega t+i\lambda y}=e^{-i{a\over Q}(3t+y)}$. At large $r$ we
then find from the
wave equation (\ref{scalareq}) the behavior \cite{lmhot}
\be
s\sim {1\over r^{3\over 2}}e^{-i {a\over Q}
(3t+y)}\cos[2\sqrt{2}{a\over Q}r+\mbox{const}]{ Y}^{(l)}
\label{scalarinf}
\ee
   The reason for the slowness of the falloff at large $r$ is the
following. Since
   $\omega>|\lambda|$, we find that at large $r$  not all the energy in
the perturbation
is tied to the $S^1$ momentum, and the residual energy goes to radial
motion; this causes the perturbation to
leak away to asymptotic infinity at late times. Normalizability at
infinity is thus seen to require
\be
\omega=|\lambda|
\label{ppone}
\ee
If we impose (\ref{ppone})
        on the solution for $s$, then we see that the solution regular
at $r=0$ is
\be
s\sim  (r^2+a^2)^l Y^{(l)}
\ee
For the choice (\ref{ppone}) there are two solutions in the outer
region with radial dependences
\be
(i)~s\sim r^{-(2l+2)}, ~~(ii)~s\sim r^{2l}
\ee
but the inner region solution  matches onto the {\it growing}
solution (ii) of the outer region, and
we again get no normalizable solution.\footnote{For $\omega=|\lambda|$
the scalar equation (\ref{scalareq})
can be exactly solved in terms of hypergeometric functions, and the
non-existence of a normalizable solution
can be explicitly seen.}

Thus we see that it is quite nontrivial that for the $(B,w)$ system of
fields the normalizable solutions
of the inner and outer regions match up at leading order. We will now
proceed to check the matching to
higher orders in $\epsilon$.

\newsection{Matching at the next order ($O(\epsilon)$)}

We wish to develop a general perturbation scheme that will
correct our solution to higher orders in $\epsilon$. It turns out that
the inner region solution does not get corrected in a nontrivial way at
order $\epsilon$. In this section we first explain the general scheme,
then apply it to
    the  outer region to get the $O(\epsilon)$ corrections,
and then explain how to match these to the inner region solution so
that the entire solution is established
to $O(\epsilon)$.

\subsection{The perturbation scheme}

The `outer region' of our geometry $r>>a$ is described  to
leading order by the metric (\ref{outer}). We must now take into
account the corrections that arise because  the exact geometry
(\ref{el}) departs from this leading order form. In particular we
get small `off-diagonal' components $g_{\mu a}$ in the metric  and
also small components  like $\bar H_{\mu\nu a},
\bar H_{\mu a b}$ of $\bar H_{ABC}$. We develop a systematic way to
handle these
corrections so that we will get the full solution as a series in
$\epsilon$.

We expand the background and perturbations as
\begin{eqnarray}
g_{MN}&=& g^{0}_{MN}+  g^{1}_{MN}\label{qfirst}\\
H & =& H_{0} +  H_{1}\\
F &=& F_{0}+  F_{1}\\
* &=& *_{0} +  *_{1} \\
w &=& w_{0}+  w_{1} \\
\nabla^{2} & = & \nabla_{0}^{2} + \nabla_{1}^{2}\label{qlast}
\end{eqnarray}
The metric $g^0_{MN}$ is the metric (\ref{outer}) we had written
earlier for the outer region.  To get $g^1_{MN}$ we take the
difference between the full metric (\ref{el}) and the outer region
metric (\ref{outer}); since we are seeking only the order $\epsilon$
corrections at this stage we keep terms of order ${a\over r},
{a\over \sqrt{Q}}$ in $g^1$ and discard higher order corrections.
Similarly we obtain $\bar H^1$. The operation $*_0$ is  defined
using the metric $g^0$, and $*_1$ contains the corrections needed
to give the $*$ operation in the full metric (upto the desired
order of approximation). $\nabla_0^2$ is the Laplacian on the metric
$g^0$
and $\nabla_1^2$ corrects this (to the desired accuracy) to the
Laplacian on the full metric.

To illustrate the general approximation scheme it is convenient to
write the perturbation
equation  (\ref{eom1}) in form language
\be F+ * F + w
\bar{H}=0 \label{formeom}
\ee
Inserting the  expansions (\ref{qfirst})-(\ref{qlast}) in
(\ref{formeom}),(\ref{eom2}) we
get
\begin{eqnarray}
F_{0} + *_{0} F_{0} + w_{0} \bar{H}_{0} &= & 0\nonumber \\
\nabla_{0}^{2} w_{0} -\frac{1}{3} \bar{H}_{0}^{MNP} F_{0MNP} &=&0
\label{firstset}\\
F_{1} + *_{0} F_{1} + w_{1} \bar{H}_{0} &=& S \nonumber \\
\nabla_{0}^{2} w_{1} - \frac{1}{3} \bar{H}_{0}^{MNP} F_{1MNP} &=&
S_{w}
\label{firstordereqns}
\end{eqnarray}
where $S_{w}$ and $S$  are defined by
\begin{eqnarray}
S &=& - w_{0}\bar{H}_{1} - *_{1} F_{0} \nonumber \\
S_{w} &=& -\nabla_{1}^{2} w_{0} + \frac{ 1}{3} \bar{H}_{1}^{MNP}
F_{0MNP}
\label{defnofsource}
\end{eqnarray}
Eqs.(\ref{firstset}) are just the leading
order
equations that give the leading order solution found  for the outer
regions in the last section.  Eqs.(\ref{firstordereqns})
    give the first order corrections. Note that the LHS of
these equations have the same form as the leading order equations, so
we need to solve the same equations again but this time with
source terms $S,S_{w}$. These source terms can be explicitly calculated
from the background geometry and the leading order solution.

\subsection{Expanding in spherical harmonics}

Even though the problem does not have exact spherical symmetry,
it is convenient to decompose fields into spherical harmonics on
$S^3$. The breaking of spherical symmetry is then manifested by
the fact that higher order corrections to the leading order
solution contain spherical harmonics that differ from the
harmonic chosen at leading order.  We write
\begin{eqnarray} w & = & e^{-i \frac{a}{Q} u } \tilde{w}^{I_{1}}
Y^{I_{1}}
\\ B_{\mu\nu} & =& e^{-i\frac{a}{Q}u} b_{\mu\nu}^{I_{1}}Y^{I_{1}} \\
B_{\mu a} & =& e^{-i\frac{a}{Q}u} b_{\mu}^{I_{3}}Y^{I_{3}}_{a} \\
B_{ab} & =& e^{-i\frac{a}{Q}u}
b^{I_{1}}\epsilon_{abc}\partial^{c}Y^{I_{1}}
\end{eqnarray}

The $Y^{I_{1}}$ are  normalized scalar spherical harmonics on the
unit 3-sphere. Their orders can be described by writing the
rotation group of $S^3$ as $so(4)=su(2)\times su(2)$. The
$Y^{I_1}$ are representations $(l,l)$ of $su (2)\times su(2)$,
with $l=0, {1\over 2}, 1, \dots$. These harmonics satisfy \be
\nabla^{2} Y^{I_{1}} = - C(I_{1}) Y^{I_{1}}, ~~ C(I_{1}) = 4
l(l+1) \ee
\be \nabla_{[a}\nabla_{b]}Y^{I_{1}} =0 \ee

         The $Y^{I_{3}}_{a}$ are  normalized vector spherical harmonics.
They fall into two classes, one with $su(2)\times su(2)$
representations $(l,l+1)$ and the other with $(l+1,l)$. (Again
$l=0, {1\over 2}, 1, \dots$.) We have
\begin{eqnarray}
\nabla^{a}Y^{I_{3}}_{a} & = & 0 \\
\nabla_{a} Y_{b}^{I_{3}} - \nabla_{b} Y_{a}^{I_{3}} &=&
\zeta(I_{3}) \epsilon_{abc} Y^{I_{3} c}
\end{eqnarray}
where
\begin{equation}
\zeta(I_{3}) = \left\{ \begin{array}{ll}  -2 (l+1), & I_{3} = (l+1,l) \\
\ 2 (l+1),  & I_{3} = (l,l+1) \end{array} \right.
\end{equation}

More details on spherical harmonics are given in Appendix \textbf{A}.

\subsection{Outer region: Solving for the first order corrections}
\label{firstordersubsection}

Returning to the field equations (\ref{firstordereqns}), we compute the
sources $S$,
finding
\begin{eqnarray}
S_{tr\theta} &=& \frac{Q}{2(Q+r^2)^2}\frac{1}{r^{2\ell+1}}
\partial_{\psi}Y^{I_{1}} \tan\theta e^{-i\frac{a}{Q}u} \nonumber \\
S_{tr\psi} &=& - \frac{Q}{(Q+r^2)^2}
\frac{1}{2r^{2\ell+1}} \left[ \sin\theta\cos\theta
\partial_{\theta}Y^{I_{1}} + 2\frac{(\ell+3)r^2 +
(\ell+1)Q }{Q+r^2} Y^{I_{1}} \cos^2 \theta \right]e^{-i\frac{a}{Q}u}
\label{jtrpsi} \nonumber \\
S_{yr\theta} &=&  - \frac{Q}{2(Q+r^2)^2}\frac{1}{r^{2\ell+1}}
\partial_{\phi}Y^{I_{1}} \cot\theta e^{-i\frac{a}{Q}u} \nonumber \\
S_{yr\phi} &=&  \frac{Q}{(Q+r^2)^2} \frac{1}{2r^{2\ell+1}}
\left[ \sin\theta\cos\theta \partial_{\theta}Y^{I_{1}} -
2\frac{(\ell+3)r^2 + (\ell+1)Q }{Q+r^2} Y^{I_{1}} \sin^2 \theta
\right]e^{-i\frac{a}{Q}u}\nonumber  \\ \label{jyrphi}
\end{eqnarray}
The source $S_w$ is zero at this order.

We can decompose these sources into scalar and vector spherical
harmonics
\be S_{\mu\nu a} = s_{\mu\nu}^{I_{3}} Y^{I_{3}}_{a} +
t_{\mu\nu} ^{I_{1}}
\partial_{a}Y^{I_{1}} \label{sourcedec}
\ee

Substituting this decomposition in (\ref{firstordereqns}) we get the
equations
\begin{eqnarray}
b_{1\mu\nu}^{I_{1}}
-\frac{r}{Q+r^2}\tilde{\epsilon}_{\mu\nu\lambda}
\partial^{\lambda}b^{I_{1}}_{1} &=&t_{\mu\nu}^{I_{1}}
\label{a1}\label{jone} \\
\partial_{t}b^{I_{3}}_{1y}-\partial_{y} b^{I_{3}}_{1t} + \zeta(I_{3})
\frac{r^3}{(Q+r^2)^2} b^{I_{3}}_{1r} &= & 0\label{first}
\label{jthree}\\
\partial_{r}b^{I_{3}}_{1t}-\partial_{t} b^{I_{3}}_{1r} + \zeta(I_{3})
\frac{b^{I_{3}}_{1y}}{r} &= & s^{I_{3}}_{tr}
\label{second}\label{jfour}\\
\partial_{y}b^{I_{3}}_{1r}-\partial_{r} b^{I_{3}}_{1y} - \zeta(I_{3})
\frac{b^{I_{3}}_{1t}}{r} &= & s^{I_{3}}_{ry} \label{third}
\end{eqnarray}
\begin{eqnarray}
\partial_{r}\left( \frac{r^3}{(Q+r^2)^2}\partial_{r}b^{I_{1}}_{1}\right)
+\frac{r}{(Q+r^2)^2}\left[ 2 Q \tilde{w}^{I_{1}}_{1} - C(I_{1})
b^{I_{1}}_{1}\right] &=& 0~~~ \label{jtwo}\\
\frac{1}{r(Q+r^2)}\partial_{r}\left(
r^3\partial_{r}\tilde{w}^{I_{1}}_{1}\right)
-\frac{C(I_{1})}{(Q+r^2)}\tilde{w}^{I_{1}}_{1}-
\frac{8Q}{(Q+r^2)^3}\left[ Q \tilde{w}^{I_{1}}_{1}-
C(I_{1})b^{I_{1}}_{1}\right]&=&0~~~~~~~~\label{jfive}
\end{eqnarray}

Eq.(\ref{jone}) yields $b^{I_1}_{\mu\nu}$ once we know $b^{I_1}$; the
source components $t^{I_1}_{\mu\nu}$
are listed in Appendix B. Eqs.(\ref{jtwo}),(\ref{jfive})
allow the
trivial solution
\be
b_{1} = \tilde{w}_{1} =0
\ee
which we adopt, since other solutions would just amount to shifting the
leading order solution taken for $b, w$.
Eq.(\ref{jthree}) yields $b_r=0$. Eqns.(\ref{jfour}),
(\ref{third}) are nontrivial and yield the solution ($u=t+y,~ v=t-y$)
\begin{eqnarray}
b_{1ua} =& b_{1u}^{I_{3}}Y^{I_{3}}_{a}&= \frac{i a}{2} \sqrt{\frac{
\ell}{(2\ell+1)
(\ell+1)}}\frac{Q}{r^{2\ell}(Q+r^2)^2}Y^{(\z+1,\z)}_{a}   +\ \ \ \
\  \ \ \ \ \ \\
        & & -\frac{ia}{4r^{2l}} \left(\sqrt{\frac{2\ell-1}{\ell(2\ell+1)}
}\frac{ Q}{(Q+r^2)^2}-\frac{1}{Q }\sqrt{\frac{4l^2-1}{l^3}} \right)
Y^{(\z-1,\z)}_{a} \\
b_{1va} &=&
\frac{ia}{4}\sqrt{\frac{1}{(\ell+1)}}\frac{Q}{r^{2\ell}(Q+r^2)^2}Y^{(\z,
\z+1)}_{a}
\end{eqnarray}

\subsection{Matching at order $\epsilon$}

\subsubsection{The inner region solution to order $\epsilon$}

Above we have applied the general scheme (\ref{firstordereqns}) to
find the outer
region solution to order $\epsilon$.
In general we would have to apply a similar scheme to correct the inner
region solution as well. But it turns out
that the expansion in the inner region goes in powers of $\epsilon^2$.
Since at this stage we are only matching terms
of order $\epsilon^0, \epsilon^1$ we do not need to perform any extra
computation for the inner region, and the solution
(\ref{first1})-(\ref{last1}) is already
correct to the desired order. But to effect the comparison with the
outer region we perform two manipulations on the  inner region
solution. First we
express the set $B_{ta}=\{ B_{t\theta}, B_{t\psi}, B_{t\phi} \}$ and 
the set
$B_{ya}$ in terms of scalar and vector harmonics
\begin{displaymath}
B_{ta} = \frac{i a e^{-i \frac{a}{Q} u} }{2Q (r^2+a^2)^{l} } \left[
\frac{\sqrt{l} Y^{(l+1,l)}_{a} }{\sqrt{(2l+1)(l+1)}} +
\frac{Y^{(l,l+1)}_{a}}{2
\sqrt{l+1}}+\ \ \ \ \ \ \  \ \ \ \  \ \ \ \ \ \  \ \ \ \ \ \ \ \ \ \ \
\  \ \ \ \ \ \ \  \right. \end{displaymath}\begin{displaymath}\left.\ \
\ \ \ \ \ \ \ \ \ \ \ \ \   \frac{l+1}{2l} \sqrt{\frac{2l-1}{l(2l+1)} }
Y^{(l-1,l)}_{a}  + \frac{\partial_{a} Y^{(l)}}{4 l^2 (l+1)}
\right]\end{displaymath} 
\begin{displaymath}
B_{ya} = \frac{i a e^{-i \frac{a}{Q} u} }{2Q (r^2+a^2)^{l} } \left[
\frac{\sqrt{l} Y^{(l+1,l)}_{a} }{\sqrt{(2l+1)(l+1)}} -
\frac{Y^{(l,l+1)}_{a}}{2
\sqrt{l+1}}+\ \ \ \ \ \ \  \ \ \ \  \ \ \ \ \ \  \ \ \ \ \ \ \ \ \ \ \
\  \ \ \ \ \ \ \  \right. \end{displaymath}\begin{equation}\left.\ \ \
\ \ \ \ \ \ \ \ \ \ \ \   \frac{l+1}{2l} \sqrt{\frac{2l-1}{l(2l+1)} }
Y^{(l-1,l)}_{a}  - \frac{(2l-1)\partial_{a} Y^{(l)}}{4 l^2 (l+1)}
\right]\label{jten}\end{equation}

Next we perform a gauge transformation on $B_{MN}$
\be
B_{MN} \rightarrow B_{MN} + \nabla_{M} \Lambda_{N} - \nabla_{N}
\Lambda_{M}
\ee
Choosing
\begin{eqnarray}
\Lambda_{t} &=& \frac{i }{8 l^2 (l+1)}\frac{a }{Q (r^2+a^2)^{l}
}Y^{(l)} e^{-i \frac{a}{Q} u}\\
\Lambda_{y} &=& -\frac{i(2l-1) }{8 l^2 (l+1)}\frac{a }{Q (r^2+a^2)^{l}
}Y^{(l)} e^{-i \frac{a}{Q} u}
\end{eqnarray}
we remove the components proportional to $\partial_{a} Y^{(l)}$ in
(\ref{jten}), while getting additional terms
in other components of $B$. In particular
\bea
B_{tr} & =& \frac{i }{4 l (l+1)}\frac{a r }{Q (r^2+a^2)^{l+1} }Y^{(l)}
e^{-i \frac{a}{Q} u} \approx \frac{i }{4 l (l+1)}\frac{a }{Q r^{2l+1}
}Y^{(l)} e^{-i \frac{a}{Q} u}\\
B_{yr} & =& \frac{i(2l^2+ 1) }{4 l (l+1)}\frac{ar }{Q (r^2+a^2)^{l+1}
}Y^{(l)} e^{-i \frac{a}{Q} u}\approx \frac{i(2l^2+ 1) }{4 l
(l+1)}\frac{a }{Q r^{2l+1} }Y^{(l)} e^{-i \frac{a}{Q} u}
\eea
We will see that with this gauge choice we will get a direct agreement
of $B_{MN}$ between the outer and inner regions.

\subsubsection{The outer region solution to order $\epsilon$}

We had solved the field equations to first order in $\epsilon$ for the
outer region in subsection (\ref{firstordersubsection}) above. We list
the complete solution
thus obtained to order $\epsilon$

\begin{eqnarray}
w&=& \frac{e^{-i\frac{a}{Q} u} }{r^{2\z}(Q+r^2) } Y^{(\z)} \nonumber
\\
B_{\theta\psi} &=& \frac{1}{4 l} \frac{ e^{-i\frac{a}{Q} u} }{r^{2l}}
\cot\theta \partial_{\phi}Y^{(l)} \nonumber    \\
B_{\theta\phi} &=& -\frac{1}{4 l}\frac{e^{-i\frac{a}{Q} u}}{r^{2l}}
\tan\theta \partial_{\psi}Y^{(l)} \nonumber    \\
B_{\psi\phi} &=& \frac{1}{4 l}\frac{e^{-i\frac{a}{Q} u}}{r^{2l}}
\sin\theta\cos\theta \partial_{\theta}Y^{(l)}\nonumber \\
B_{ty} & = & - \frac{1  }{2 (Q+r^2)^{2} } \frac{ e^{-i \frac{a}{Q} u}
}{r^{2l-2} } Y^{(l)}\nonumber     \\
B_{tr} &=&-\frac{i a }{r^{2l+1} } \left( \frac{ Q }{ (Q+r^2)^3}
\frac{\left[(\ell+2) r^2 +\ell Q\right]}{4\ell(\ell+1)} -
\frac{1}{4 l Q  }\right) Y^{(\z)}e^{-i \frac{a}{Q} u}\nonumber    \\
B_{yr} &=&  \frac{i  a}{r^{2l+1}} \left(\frac{(2\ell-1)  Q  }{
(Q+r^2)^3}  \frac{\left[(\ell+2) r^2 +\ell Q\right]}{4\ell(\ell+1)} +
\frac{1}{4\ell Q  }\right) Y^{(\z)}e^{-i\frac{a}{Q}u}\nonumber    \\
B_{ta}&=& \frac{i a Q e^{-i \frac{a}{Q} u}  }{2r^{2l} (Q+r^2)^{2} }
\left[ \sqrt{\frac{l}{(2l+1)(l+1)}} Y^{(l+1,l)}_{a} +
\frac{Y^{(l,l+1)}_{a}}{2 \sqrt{l+1}} - \frac{1}{2}
\sqrt{\frac{2l-1}{l(2l+1)} } Y^{(l-1,l)}_{a}  \right]\nonumber    \\
        &&~~~~~~~~~~~~~~~~~~~~~~+ \frac{i a}{4 Q r^{2l}
}\sqrt{\frac{4l^2-1}{l^3}} Y_{a}^{(l-1,l)}\nonumber    \\
B_{ya}&=& \frac{i a Q e^{-i \frac{a}{Q} u}  }{2r^{2l} (Q+r^2)^{2} }
\left[ \sqrt{\frac{l}{(2l+1)(l+1)}} Y^{(l+1,l)}_{a} -
\frac{Y^{(l,l+1)}_{a}}{2 \sqrt{l+1}} - \frac{1}{2}
\sqrt{\frac{2l-1}{l(2l+1)} } Y^{(l-1,l)}_{a}  \right]\nonumber    \\
        &&~~~~~~~~~~~~~~~~~~~~~~+ \frac{i a}{4 Q r^{2l}
}\sqrt{\frac{4l^2-1}{l^3}} Y_{a}^{(l-1,l)}
\label{outersolution}
        \end{eqnarray}

\subsubsection{Comparing the inner and outer solutions at order
$\epsilon$}

In the region where we match solutions we have to substitute at the
present order of approximation
\be
\frac{1}{(r^2+a^2)^{l}} \approx \frac{1}{r^{2l} }, ~~~~
\frac{1}{(Q+r^2)} \approx \frac{1}{Q}
\ee
We then find agreement between the inner region solution (in the gauge
discussed above) and outer region solution
(\ref{outersolution}).

\newsection{Matching at higher orders}

We follow the same scheme to extend the computation to higher orders in
$\epsilon$. At each stage the
sources $S, S_w$ get contributions from all the terms found at
preceeding orders. The computations are straightforward though tedious,
and most are done using symbolic manipulation programs.

The solutions obtained for the inner region are listed in Appendix \textbf{B}.
We have given the solutions in the NS sector
coordinates; they must be spectral flowed to the R sector and gauge
transformations performed to see
directly the agreement with the outer region solutions. As mentioned
before the perturbation series in the NS sector of the inner region
proceeds in even powers of $\epsilon$, and the odd powers in $\epsilon$
result from the spectral flow (\ref{spectral}).

The solutions obtained for the outer region are listed in Appendix \textbf{C}.
These are already in R sector coordinates. Note that
at each order when we solve the equations with sources we have to
choose a homogeneous part to the solution as well, and these parts have
been chosen to give regularity everywhere as well as agreement between
the inner and outer regions.

We carry out the computation of the solution in each region
to order $O(\epsilon^3)$. We find complete agreement between the inner
and outer region solutions
upto the order investigated.
At each stage of the computation  there is the
possibility of finding
that some field
is growing at infinity, and it is very nontrivial that this does not
happen
for any field at any of the orders studied.
Thus we expect that the exact solution does exist and is likely to  be
expressible in closed form.

At all the orders that we have investigated the scalar $w$ can be seen
to arise from expansion of the solution
\be
w=\frac{e^{-i \frac{a}{Q} u} Y^{(l)}}{(r^2+a^2)^{\z}
(Q+f )}, ~~~~~f=r^2+a^2\cos^2\theta
\ee
Note that this expression involves just the combinations $(r^2+a^2), f$
which appear in the geometry
(\ref{el}).
       We do not have a similar compact expression for the $B$ field; it
is
plausible that the compact form would
       require us to express this 2-form field as part 2-form and part
6-form
(the magnetic dual representation).
       We hope to investigate this issue elsewhere.

\newsection{Discussion}

We have constructed regular, normalizable supergravity perturbations in
the inner and outer regions by a process of successive
corrections, and observed that at each order the solutions agree in the
domain of overlap. This agreement is very nontrivial, and we take this
as evidence for the existence of an exact solution to the problem --
i.e. we expect that there exists a  regular perturbation on the
2-charge D1-D5  geometry (\ref{el}) which carries one unit of momentum
charge and adds one unit of energy
(thus yielding an extremal 3-charge solution). We now return to our
initial discussion of black hole interiors, and
the significance of this solution in that context.

The usual picture of a black hole has a horizon, a singularity at the
center, and `empty space' in between.
Abstract arguments given in the introduction suggested a different
picture where the interior was
nontrivial and exhibited the degrees of freedom contributing to the
entropy. The 2-charge extremal system
turned out to look like this latter picture -- its properties
(a$'$)-(c$'$) listed in the introduction
matched the suggested properties (a)-(c).
What about the 3-charge extremal hole? This latter hole has become a
benchmark system for understanding black holes, and any lessons deduced
here likely extend to all holes in all dimensions.

The metric conventionally written for the D1-D5-P extremal system is
\bea
ds^2&=&{1\over \sqrt{(1+{Q_1\over r^2})(1+{Q_5\over
r^2})}}[-dudv+{Q_p\over r^2}dv^2]\nonumber\\
~~&+&\sqrt{(1+{Q_1\over r^2})(1+{Q_5\over
r^2})}[dr^2+r^2d\Omega_3^2]+\sqrt{(1+{Q_1\over r^2})\over (1+{Q_5\over
r^2})}dz_adz_a
\label{done}
\eea

\medskip
\begin{figure}[t]
\hspace{0.8in}
\includegraphics{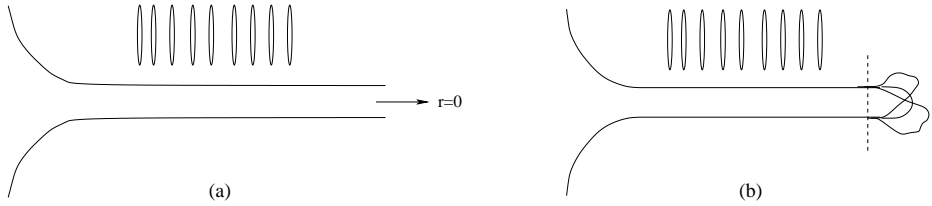}
\caption{ \small{(a) Naive geometry  for the $3$-charge extremal system.
(b) Expected structure for the system.}} \label{nfig7}
\end{figure}

This is similar to the `naive' metric (\ref{naive}) of the 2-charge
D1-D5
extremal system, except that the $y$
circle stabilizes to a fixed radius as $r\rightarrow 0$ instead of
shrinking to zero size (we picture the geometry
(\ref{done}) in Fig.\ref{nfig7}(a)). The geometry (\ref{done})
has a completion that it continues past the
horizon at $r=0$ to the `interior' of the black hole, where we have  a
timelike singularity -- the metric is just a 4+1 analogue of the
extremal Reissner-Nordstrom black hole.

In a roughly similar manner one might have asked if the 2-charge metric
continues past the `horizon' $r=0$ to another region, but here we do
know the answer -- the naive metric (\ref{naive}) is incorrect, and the
actual
geometries `cap off' before reaching $r=0$. We are therefore led to ask
if a similar situation holds for the
3-charge system, so that the actual geometries `cap off' before
reaching $r=0$ as in Fig.\ref{nfig7}(b). We would then draw the
`horizon' as a
surface which bounds the region where the geometries differ from each
other significantly; this surface is indicated by the dashed line in
Fig.\ref{nfig7}(b). Note that for the 3-charge system the area of
this `horizon' will give {\it exactly}
\be
{A\over 4G}=S_{micro}=2\pi\sqrt{n_1n_5n_p}
\label{dtwo}
\ee
This is because in the naive metric (\ref{done}) the cross sectional
area of the throat
saturates to a constant $A$ as $r\rightarrow 0$, and it is this same
value $A$ that will be  picked up at the location
of the dashed line in  Fig.\ref{nfig7}(b). But from \cite{stromvafa}
we know that
this area $A$ satisfies (\ref{dtwo}). (For the 2-charge case we could
find $A$ only upto a
factor of order unity, since the $y$ circle of the cross section was
shrinking with $r$, and the natural uncertainty in  the location  of
the `horizon surface' leads to a corresponding uncertainty in $A$.)

Thus for the 3-charge system the nontrivial issue is not horizon area
(which we see will work out anyway) but the nature of the geometry
inside the horizon.
The computation of this paper has indicated that if we have one unit of
$P$ then at least one
extremal state
\be
|\Psi\rangle=J^-_{-1}|0\rangle_R
\label{dthree}
\ee
of the 3-charge system is described by a geometry like
Fig.\ref{nfig7}(b) and not
by Fig.\ref{nfig7}(a).  It may be argued though that
the 2-charge extremal states and the state (\ref{dthree}) are not
sufficiently like generic black hole states
to enable us to conclude that Fig.\ref{nfig7}(b) is the generic
geometry of the
3-charge system. Here we give
several arguments that counter this possibility:

\bigskip

(a) {\it Is the 2-charge system like a black hole?}\quad
It is sometimes argued that the 2-charge extremal system is
not really a black hole since the horizon area  vanishes classically.
        We argue against this view. The microscopic entropy of the
2-charge
extremal system (S=$2\sqrt{2}\pi\sqrt{n_1n_5}$) arises by partitions of
$N=n_1n_5$ in a manner similar to the entropy $2\pi\sqrt{n_1n_5n_p}$ of
the 3-charge extremal system which  arises from partitions of
$N=n_1n_5n_p$.
The `horizon'  that we have constructed for the 2-charge system
satisfies
$S\approx A/4G$, so this `horizon' area is $\sim \sqrt{n_1n_5}$ times
$(l_p)^3$, and is thus {\it not} small at all in planck units.

Why then do we think of this horizon as small? The 2-charge metric has
factors like $\sim (1+{Q_1\over r^2}), (1+{Q_5\over r^2})$. Assuming
$Q_1\sim Q_5$ and $n_1\sim n_5\sim n$ we find that the geometry has a
scale,
the `charge radius', which grows with $n$ as  $r\sim Q^{1\over 2}\sim
n^{1\over 2}$. Since the horizon is a 3-dimensional surface, and we 
have found $S_{micro}\sim n\sim {A\over 4G}$, the
horizon radius is $r\sim n^{1\over 3}$.  Suppose we take the classical 
limit $n\rightarrow\infty$ and
then scale the metric so that the charge radius is order unity. In this 
limit the horizon radius will {\it vanish}. For the
3-charge system, both the charge radius and the horizon radius behave
as $r\sim n^{1\over 2}$, so the horizon radius remains nonzero in the
analogous classical
limit.

But this behavior of classical limits does not imply that the 2-charge
system has an
ignorable horizon -- the horizon does give the correct entropy, and the
presence of the other, larger, length scale appears irrelevant to the
physics inside this horizon.
The region $r\sim Q^{1\over 2}$ is far removed from the horizon
region, and simply governs the changeover from `throat geometry' to
`flat space'.

\medskip

(b) {\it Return time $\Delta t_{CFT}$:}\quad For the 2-charge system,
the naive metric is (\ref{naive}). If we throw a test particle down
the throat of this naive metric, it does not return after any finite 
time. In the dual CFT
however an excitation absorbed by the `effective string'
can be re-emitted after a time $\Delta t_{CFT}<\infty$. How do we 
resolve
this contradiction? One might think that nonperturbative effects cause
the test particle to turn back from some point in the throat of the
naive geometry,
but this cannot be the case since the return time $\Delta t_{CFT}$ is
different for different states of the 2-charge system (the length of
the components of the effective string are different for different
states). The resolution of this puzzle was that the throats were
capped; the cap was different for different states \cite{lm4}, and we
get (\ref{equal}).

The CFT for the 3-charge system is described by the same effective
string; we just have additional momentum excitations
on the effective string. We would thus again have some finite time
$\Delta t_{CFT}$ after which an excitation should be emitted back from
the system, and the requirement (\ref{equal}) then suggests that
Fig.\ref{nfig7}(b) is the correct picture for the general states of
the 3-charge
system, rather than Fig.\ref{nfig7}(a).

\medskip

(c) {\it Fractionation:}\quad We have argued that the interior of the
horizon is not the conventionally
assumed `empty space with central singularity'. How can the classical
expectation be false over such large
length scales? The key physical effect is `fractionation'. If we excite
a pair of left and right vibrations on a string of length
$L$, the minimum excitation threshold is $\Delta E={2\pi\over
L}+{2\pi\over L}={4\pi\over L}$. But if we have a bound state
of $n$ strings, then we get one long string of length $nL$, and the
threshold drops to ${4\pi\over nL}$ \cite{dmone}.
If we start with 2-charges, $n_1$ D1 branes and $n_5$ D5 branes, then
the excitations of the third charge, momentum,
come in even smaller units, and $\Delta E={4\pi\over n_1n_5 L}$
\cite{maldasuss}. If we assume more generally that for the bound state
of mutually supersymmetric branes the excitations always fractionate in
this way, then we find that
the excitations of the D1-D5-P hole are such that they extend to a
radial distance that is just the horizon scale \cite{emission}.
For the 2-charge FP where we have explicitly constructed all geometries
this fractionation effect can be directly seen --
because the momentum waves are fractionally moded on the multiply wound
F string, the strands of the F string separate
and spread over a significant transverse area, which extends all the
way to the `horizon'.

\medskip

(d) {\it Other 3-charge states:}\quad The general perturbations around
the 2-charge solution that we have chosen decompose into two classes:
The antisymmetric field $+$ scalar perturbations (which we have
analyzed) and the metric $+$ self-dual field perturbations. We have
checked upto leading order ($\epsilon^0$) that the latter class gives a
regular solution as well.
Further, the 2-charge solution that we started with may appear special
(It has for instance angular momentum
${n_1n_5\over 2}$ in each $su(2)$ factor, while the generic 2-charge
state has negligible net angular momentum) but we have also checked 
that at
leading order we get regular perturbations for all starting 2-charge
geometries. In principle all these computations could be carried out to
higher orders in $\epsilon$, but the technical complexities would be
greater due to less symmetry in the starting configuration.

One might think that if we increase the the momentum $p$ then we might 
get a horizon.
For $p=1$ we have seen that the perturbation is smooth, so there is no 
hint of an incipient horizon.  Suppose for some $p=p_0$ a horizon
just about forms; this horizon will be of radius zero at $p=p_0$, and 
larger at larger $p$. But what will be the
location of the horizon at $p=p_0$? There is no special point in the 
starting 2-charge geometries; they
are just smoothly capped throats. It thus appears more likely that 
adding momentum will just
give more and more complicated configurations, but with no 
special point which could play the role of a singularity.

\medskip

(e) {\it Nonextremal holes:}\quad Having found the above structure for
extremal systems, we expect a similar structure
for near extremal and also neutral holes, with the difference that the
branes in the extremal systems are replaced
by a collection of branes and anti-branes. Indeed, for the non-extremal
D1-D5-P system it is known that the entropy
of holes arbitrarily far from extremality can be reproduced exactly if
we assume that the energy is
optimally partitioned between branes and anti-branes while reproducing
the overall charges and mass \cite{hms}.

In an interesting recent paper \cite{emparan} it was argued that the
`black ring' solutions carrying D1-D5-P charges
(plus nonextremality) had pathologies like closed timelike curves and
thus it was not possible to add momentum by boosting to
general rotating D1-D5 states. It was observed however that it might be 
possible to add momentum in other ways to get
a 3-charge state.
Our construction {\it
does} take a D1-D5 state with some angular momentum, and adds one unit
of momentum. But looking at the form of the perturbation it can be seen 
that the momentum was not obtained by a boost.

More generally, generating metrics by boosting a `naive' nonextremal geometry 
will {\it not} give the correct states
of the system.   In such a construction  we start with a nonextremal 
black  hole or black ring geometry, where the
metric in the interior of the horizon is just the
classically expected one (similar in spirit to Fig.1(a) for a black
hole).
But we have argued that such an interior metric is {\it not} a correct
description for the
region inside the horizon; this region
we believe is very nontrivial,  with details that necessarily depend on
the particular state which the system takes (out of the $e^S$ possible
states).\footnote{One should
not use the `correspondence principle'
\cite{hp} to obtain a qualitative understanding of what might happen
inside horizons. It was
shown in \cite{emission} that at coupling $g<g_c$ the energy added to a
string goes to
exciting vibrations, while at $g>g_c$ the energy goes to creating {\it
brane-antibrane pairs}.
(Here $g_c$ is the coupling at the `correspondence point' where the
string turns to  a black hole.) It is these
brane-antibrane pairs that have the small energy gaps and large phase
space to `fill up' the
interior of the horizon. } Instead one should start with one of the 
`correct' states for system, and then construct
the possible deformations that add momentum.

We can emphasize this point in another way, using just the 2-charge 
system.
Suppose we start with the {\it naive}
metric for the {\it nonextremal}
F string. This metric will have cylindrical symmetry around the axis of
the F string. We can boost and add momentum, still
keeping the cylindrical symmetry and getting F and P charges. We can
then  take the non-extremality to zero. This process will
      reproduce the
{\it naive} metric (\ref{naivefp}) of the extremal FP system. To get
the {\it correct} metrics for  extremal FP  starting
from  non-extremal FP  we would have to start with one of the correct
interior states for the {\it nonextremal} FP system.

\bigskip

Clearly what we need next is a construction of the generic 3-charge
configuration (i.e. with the P charge not small).
  It is important that the solutions
represent true
       bound states rather than just multi-center brane solutions that
are classically supersymmetric. (Some families of metrics with 3 charges
have been constructed before
(e.g. \cite{myers}) but we are not aware of any set that actually 
describes
the bound states that we wish to study.)\footnote{We thank D. Mateos and
O. Lunin for discussions on this point.}  It is possible that the
generic state is not well approximated
by a classical configuration; what we do expect though on the basis of
all that was said above is that the region where the different states
depart from each other will be of the order the horizon size and not
just a planck sized region near the singularity.

\section*{Acknowledgements}

This work was supported in part by DOE grant DE-FG02-91ER-40690. We are
grateful to Oleg Lunin
for many helpful discussions, and for several contributions in the
initial stages of this project. We also thank
Jan de Boer, Sumit Das, Steven Giddings, Per Kraus, David Kutasov,
Emil Martinec  and David Mateos for useful comments and
discussions.

\section*{Appendix A: Spherical Harmonics on $S^{3}$}
\begin{appendix}
\renewcommand{\theequation}{A.\arabic{equation}}
\setcounter{equation}{0}
\renewcommand{\thesubsection}{A.\arabic{subsection}}
\setcounter{subsection}{0}

In this Appendix we list the explicit forms of the various spherical
harmonics encountered in the solutions. The metric on the unit 3-sphere
is
\be
ds^2=d\theta^2+\cos^2\theta d\psi^2+\sin^2\theta d\phi^2
\ee
The harmonics will be orthonormal
\bea
\int d\Omega ~(Y^{I_1} )^*Y^{I'_1}&=&\delta^{I_1, I'_1}\nonumber\\
\int d\Omega~(Y^{I_3}_a )^*Y^{I'_3 a}&=&\delta^{I_3, I'_3}
\eea
In order to simplify notation
we have used the following abbreviations

\begin{eqnarray}
\hat{Y}^{(l)} &\equiv& Y^{(l,l)}_{(l,l)} \\
Y^{(l)} &\equiv& Y^{(l,l)}_{(l-1,l)} \\
Y^{(l+1)} & \equiv & Y^{(l+1,l+1)}_{(l-1,l)} \\
Y_{a}^{(l+1,l)} & \equiv & Y^{(l+1,l)}_{a(l-1,l)} \\
Y_{a}^{(l,l+1)} & \equiv & Y^{(l,l+1)}_{a(l-1,l)} \\
Y_{a}^{(l-1,l)} & \equiv & Y^{(l-1,l)}_{a(l-1,l)} \\
Y_{a}^{(l+2,l+1)} & \equiv & Y^{(l+2,l+1)}_{a(l-1,l)} \\
Y_{a}^{(l+1,l+2)} & \equiv & Y^{(l+1,l+2)}_{a(l-1,l)}
\end{eqnarray}

\subsection{Scalar Harmonics}

The scalar harmonics we use are (in explicit form)

\begin{eqnarray}
\hat{Y}^{(l)} & =& \sqrt{\frac{2l+1}{2}} \frac{e^{-2i l \phi}
}{\pi} \sin^{2l}\theta \\
Y^{(\z)} &=& -\frac{\sqrt{\z(2\z+1)}}{\pi} e^{-i(2\z-1)\phi + i \psi}
\sin^{2\z-1} \theta \cos\theta \\
Y^{(\z+1)} &=& \frac{\sqrt{(2\z+1)(2\z+3)}}{2\pi} e^{-i(2\z-1)\phi +
i\psi} ((\z-1)+(\z+1) \cos2\theta) \sin^{2\z-1}\theta \cos\theta
\nonumber \\
\end{eqnarray}

\subsection{Vector  Harmonics}

The vector harmonics  are given by

\begin{eqnarray}
Y^{(\z+1,\z)}_{\theta } &=& -\frac{ e^{-i(2\z-1)\phi +i \psi} }{4\pi}
\frac{\sin^{2\z-2}\theta}{\sqrt{\z+1}} \left( (2\z^2-\z+1) +
(\z-1)(2\z+1) \cos2\theta\right)\\
Y^{(\z+1,\z)}_{\psi}&=&  i \frac{ e^{-i(2\z-1)\phi +i \psi} }{4\pi}
\frac{\sin^{2\z-1}\theta \cos\theta   }{\sqrt{\z+1}} \left(
(2\z^2+3\z-1) + (\z+1)(2\z+1) \cos2\theta\right) ~~~ \\
Y^{(\z+1,\z)}_{\phi}&=& -i \frac{ e^{-i(2\z-1)\phi + i \psi} }{4\pi}
\frac{\sin^{2\z-1}\theta \cos\theta   }{\sqrt{\z+1}} \left(
(2\z^2-5\z-1) + (2\z^2+3\z+1) \cos2\theta\right) ~~~~~~~~
\end{eqnarray}
\begin{eqnarray}
Y^{(\z,\z+1)}_{\theta}&=& - \frac{ e^{-i(2\z-1)\phi +i \psi}
}{4\pi}\sqrt{\frac{4\z(2\z+1)}{\z+1}} \sin^{2\z-2}\theta \left( (\z-1)
+ \z \cos2\theta\right) \\
Y^{(\z,\z+1)}_{\psi}&=&  i  \frac{ e^{-i(2\z-1)\phi +i \psi}
}{4\pi}\sqrt{\frac{4\z(2\z+1)}{\z+1}} \sin^{2\z-1}\theta\cos\theta
\left( \z + (\z+1) \cos2\theta\right) \\
Y^{(\z,\z+1)}_{\phi}&=&  i \frac{ e^{-i(2\z-1)\phi +i \psi}
}{4\pi}\sqrt{\frac{4\z(2\z+1)}{\z+1}} \sin^{2\z-1}\theta\cos\theta
\left( (\z+2) + (\z+1) \cos2\theta\right)~~~~~~
\end{eqnarray}
\begin{eqnarray}
Y^{(\z-1,\z)}_{\theta}&=&  \frac{ e^{-i(2\z-1)\phi +i \psi}
}{2\pi}\sqrt{2\z-1} \sin^{2\z-2}\theta  \\
Y^{(\z-1,\z)}_{\psi}&=& -i \frac{ e^{-i(2\z-1)\phi + i \psi}
}{2\pi}\sqrt{2\z-1} \sin^{2\z-1}\theta\cos\theta \\
Y^{(\z-1,\z)}_{\phi}&=& -i \frac{ e^{-i(2\z-1)\phi +i \psi}
}{2\pi}\sqrt{2\z-1} \sin^{2\z-1}\theta\cos\theta
\end{eqnarray}
\begin{eqnarray}
Y^{(\z+2,\z+1)}_{\theta}&=&  -\frac{ e^{-i(2\z-1)\phi +i \psi}
}{8\pi}\sqrt{\frac{3}{\z+2}} \sin^{2\z-2}\theta
\left[(\z-1)(2\z^{2}+\z+1) \right. \nonumber \\
& & \left. + \frac{2(4\z^{3}-\z +3)\cos2\theta}{3} +
\frac{(\z-1)(\z+1)(2\z+3)\cos4\theta}{3}\right] \\
Y^{(\z+2,\z +1)}_{\psi}&=& -i \frac{ e^{-i(2\z-1)\phi + i \psi}
}{4\pi}\sqrt{\frac{3}{\z+2}} \sin^{2\z-1}\theta\cos\theta
\left[\frac{\z(2\z^{2} +5\z -1)}{2} \right. \nonumber \\  & & \left.+
\frac{1}{3}(\z+1)(4\z^{2}+8\z -3)\cos2\theta
       + \frac{(\z+1)(\z+2)(2\z+3)}{6}\cos4\theta \right] ~~~~~\\
Y^{(\z+2,\z +1)}_{\phi}&=& i \frac{ e^{-i(2\z-1)\phi +i \psi}
}{4\pi}\sqrt{\frac{3}{\z+2}} \sin^{2\z-1}\theta\cos\theta
\left[\frac{(2\z^{3} -3\z^{2} + 3\z +4)}{2} \right. \nonumber \\ & &
\left.+ \frac{1}{3}(4\z^{3}-13\z -9)\cos2\theta  +
\frac{(\z+1)(\z+2)(2\z+3)}{6}\cos4\theta \right]
\end{eqnarray}

\section*{Appendix B: Solution -- inner region}

\renewcommand{\theequation}{B.\arabic{equation}}
\setcounter{equation}{0}
\renewcommand{\thesubsection}{B.\arabic{subsection}}
\setcounter{subsection}{0}

The supergravity equations are expressed in terms of the fields
$B_{MN}$ and $w$. It is convenient to divide the $B_{MN}$ into three
classes -- $B_{a b}, B_{\mu a}$ and $B_{\mu\nu}$ where $B_{ab}$ is an
antisymmetric tensor on $S^{3}$, $B_{\mu a}$ is a vector on $S^3$ and
$B_{\mu\nu}$ is a scalar on $S^{3}$. At a given order $\epsilon^{n}$,
the corrections to $B_{a b}$ and $B_{\mu\nu}$ at that order can be
expressed in terms of a single scalar field $b$ and the antisymmetric
tensor $t_{\mu\nu}$:
\bea
B_{ab}& =& \eps_{ a b c}e^{-2 i \z \frac{a}{Q} t }  \partial^{c} b\\
B_{\mu\nu} &=&   \frac{r}{Q}
\tilde{\epsilon}_{\mu\nu\lambda}\partial^{\lambda} \left( e^{-2 i 
\z\frac{a}{Q} t } b \right) + e^{-2 i \z
\frac{a}{Q} t } t_{\mu\nu}
\eea
Here $\epsilon_{abc}$ is the usual Levi-Civita tensor on the unit $S^3$
(with $\epsilon_{\theta\psi\phi}=\sqrt{g}$), while
$\tilde{\epsilon}_{\mu\nu\lambda}$ is the Levi-Civita tensor {\it 
density} on the
$t,y,r$ part of the metric (\ref{innerns}); thus  $\tilde{\eps}_{tyr}
=1$.
Below we will give the values of $b$ and $t_{\mu\nu}$ at each order in
the perturbation. The
1-forms $B_{t a}, B_{ya}$ and $B_{ra}$ will be given explicitly. To
avoid cumbersome notation we do not
put labels on the fields indicating the order of perturbation; rather
we list the order of all  fields in the subsection heading.

In this Appendix the solutions are in the $NS$ sector coordinates. In
order to compare with the outside we need to spectral flow these to the
R sector using the coordinate transformation
\be
\psi_{NS}=\psi-{a\over Q}y, ~~~\phi_{NS}=\phi-{a\over Q}t
\label{spectralapp}
\ee
The perturbation expansion in the NS sector coordinates has only even
powers of $\epsilon$. The spectral flow
(\ref{spectralapp}) to R sector coordinates generates odd powers in
$\epsilon$. Thus the $O(\epsilon^0)$ NS sector
computation gives $O(\epsilon^0), O(\epsilon^1)$ in the R coordinates.

The solution to a given order $\epsilon^n$ is given by the sum of the
corrections at all orders $\le n$.

\subsection{Leading Order ($O(\epsilon^0)\rightarrow O(\epsilon^0),
O(\epsilon^1)$)}

\begin{eqnarray}
b  &=& \frac{1}{4\z} \frac{ 1}{(r^2+a^2)^{\z} } Y^{(\z)}_{NS}\\
w  &=&   \frac{ 1}{Q (r^2+a^2)^{\z} } Y^{(\z)}_{NS} e^{-2 i \z
\frac{a}{Q} t } \\
B_{ta} &=& B_{ya}=B_{ra} =0 \\
t_{\mu\nu} & = & 0
\end{eqnarray}

\subsection{Second Order ($O(\epsilon^2)\rightarrow O(\epsilon^2),
O(\epsilon^3)$)}

\begin{eqnarray}
b &=& \frac{a^{2} }{ Q (r^2+a^2)^{\z}} \left[ \frac{(3\z-1) -
2\z(\z+1) \cos^2\theta }{4\z(\z+1)^2 } \right]Y^{(\z)}_{NS} \\
w &=& - \frac{ 1}{Q (r^2+a^2)^{\z} } \frac{r^{2} + a^2 \cos^2\theta
}{Q} Y^{(\z)}_{NS}e^{-2 i \z \frac{a}{Q} t } \\
B_{ta} & =& - \frac{i a}{Q^{2} (r^2+a^2)^{\z} }
\left[\left( \sqrt{\frac{\z}{(\z+1)^{5}(2\z+1)} }\right)
\left[ (2\z+1) a^2+ (\z+1)^2 r^2 \right] (Y^{(\z+1,\z)}_{a})_{NS}\right.
\nonumber \\
& &     +\left(\frac{1}{2(\z+1)^2} \sqrt{\frac{1}{(\z+1)}
}\right)  \left[ (3\z+1) a^2+ (\z+1)^2 r^2 \right]
(Y^{(\z,\z+1)}_{a})_{NS}
\nonumber \\
& & - \left. \left( \frac{1}{4\z} \sqrt{\frac{ 2\z-1}{\z(2\z+1) } }
\right)
\left[a^2+2\z r^2\right] (Y^{(\z-1,\z)}_{a})_{NS} \right] e^{-2 i \z
\frac{a}{Q} t }
\end{eqnarray}
\begin{eqnarray}
       B_{ya} & =&  -\frac{ia}{Q^{2} (r^2+a^2)^{\z} } \left[
\left(\sqrt{\frac{\z}{(\z+1)(2\z+1)} }\right) r^2
(Y^{(\z+1,\z)}_{a})_{NS} -
\frac{1}{2 \sqrt{\z+1}  } r^2 (Y^{(\z,\z+1)}_{a})_{NS}  \right.
\nonumber \\
& &-\left.\left( \frac{1}{4\z} \sqrt{\frac{ 2\z-1}{\z(2\z+1) } } \right)
\left[a^2+2\z r^2\right] (Y^{(\z-1,\z)}_{a})_{NS}\right] e^{-2 i \z
\frac{a}{Q} t }  \\
B_{ra} &=&   \frac{a^{2}}{Q (r^2+a^2)^{\z+1} }
\left[\left(  \sqrt{\frac{\z^{5}}{(\z+1)^{5}(2\z+1)}
}\right)r (Y^{(\z+1,\z)}_{a})_{NS} \right.
+\left(\frac{\z(\z-1)}{2(\z+1)^{\frac{5}{2}}}
\right)
r(Y^{(\z,\z+1)}_{a})_{NS} \nonumber \\
& &-\left. \left( \frac{1}{4\z} \sqrt{\frac{ 2\z-1}{\z(2\z+1) } }
\right)
\frac{1}{r }  \left[a^2+2\z r^2\right](Y^{(\z-1,\z)}_{a})_{NS} \right]
e^{-2 i \z \frac{a}{Q} t }
\end{eqnarray}
\begin{eqnarray}
t_{ty} & = &  \frac{r^{2}}{Q^{3} (r^2+a^2)^{l} } \left[ \left(\frac{
(2l+1) a^{2}  + (l+1)^{2} r^{2}  }{(l+1)^{2} } \right) Y^{(l)}_{NS}+
a^{2}
\frac{l}{(l+1)^{2} }  \sqrt{\frac{l}{(2l+3) } } Y^{(l+1)}_{NS}\right]
\nonumber \\ \\
t_{yr} & = & i\frac{ a r }{Q^{2} (r^2+a^2)^{l+1} } \left( \frac{ (l^2+
2l -1) a^{2}  - (l^2-1)(2l-1) r^2 }{2 l (l+1)^2 } \right) Y^{(l)}_{NS}
\nonumber \\  & & - i  \frac{a^{3} r }{Q^{2} (r^{2} + a^2 )^{l+1} }
\frac{l}{(l+1)^{2} }\sqrt{\frac{l}{(2l+3)} } Y^{(l+1)}_{NS}  \\
t_{tr} & = & i \frac{a r} { Q^{2}  (r^2+a^2)^{l} } \frac{l-1}{2 l (l+1)
} Y^{(l)}_{NS}~~~~~
\end{eqnarray}

\section*{Appendix C: Solution -- outer region}

\renewcommand{\theequation}{C.\arabic{equation}}
\setcounter{equation}{0}
\renewcommand{\thesubsection}{C.\arabic{subsection}}
\setcounter{subsection}{0}
    As was done for the inner region, we divide the field $B_{MN}$
into three
classes -- $B_{a b}, B_{\mu a}$ and $B_{\mu\nu}$. At a given order
$\epsilon^{n}$,
the corrections to $B_{a b}$ and $B_{\mu\nu}$ at that order can be
expressed in terms of a single scalar field $b$ and the antisymmetric
tensor $t_{\mu\nu}$:
\bea
B_{ab}&=&e^{-i \frac{a}{Q} u }  \eps_{ a b c} \partial^{c} b\nonumber\\
B_{\mu\nu} &=&  \left( \frac{r}{Q+r^2}
\tilde{\epsilon}_{\mu\nu\lambda}\partial^{\lambda} b +
t_{\mu\nu}\right)e^{-i \frac{a}{Q} u }
\eea
Again $\epsilon_{abc}$ is the Levi-Civita tensor on the unit $S^3$
while $\tilde\epsilon_{\mu\nu\lambda}$ is the Levi-Civita
tensor {\it density} on the $t,y,r$ part of the metric (\ref{outer});
thus $\tilde\epsilon_{tyr}=1$.
We give $b, t_{\mu\nu}$ at each order.  We also write
\bea
B_{\mu a} &=& e^{-i \frac{a}{Q} u } b_{\mu a}\nonumber\\
w&=& e^{-i \frac{a}{Q} u } \tilde{w}
\eea
We will give  $b_{\mu a}, \tilde w$ at each order.

The solution to a given order $\epsilon^n$ is given by the sum of the
corrections at all orders $\le n$.

\subsection{Leading  Order ($O(\epsilon^0)$)}

\begin{eqnarray}
    & b   &=\frac{1}{4\z} \frac{1}{r^{2\z} } Y^{(\z)}\\
    &\tilde{w} & = \frac{1}{r^{2\z}(Q+r^2) } Y^{(\z)}\\
& b_{ta}& =b_{ya}=b_{ra}=0\\
& t_{\mu\nu}& = 0
\end{eqnarray}

\subsection{First Order ($O(\epsilon^1)$)}

\begin{eqnarray}
&b&=\tilde{w}=0\\
&b_{ua}&=  \frac{ia}{2} \sqrt{\frac{ \ell}{(2\ell+1)
(\ell+1)}}\frac{Q}{r^{2\ell}(Q+r^2)^2}Y^{(\z+1,\z)}_{a} \nonumber \\
& & \ \ \ -\frac{ia}{4}
\frac{1}{r^{2\ell}}\sqrt{\frac{2\ell-1}{\ell(2\ell+1)}
}\frac{Q}{(Q+r^2)^2} Y^{(\z-1,\z)}_{a}+\frac{ia}{4Qr^{2l}
}\sqrt{\frac{4l^2-1}{l^3}} Y^{(\z-1,\z)}_{a} \\
& b_{va}& =
i\frac{a }{4}\sqrt{\frac{1}{(\ell+1)}}\frac{Q}{r^{2\ell}
(Q+r^2)^2}Y^{(\z,\z+1)}_{a}
\end{eqnarray}

\begin{eqnarray}
&t_{ty}& =0 \\
&t_{rt}& =  i a \left( \frac{  Q }{r^{2\ell+1} (Q+r^2)^3}
\frac{\left[(\ell+2) r^2 +\ell Q\right]}{4\ell(\ell+1)} -
\frac{ 1}{4\ell Q r^{2\ell+1} }\right) Y^{(\z)}\\
&t_{yr}&=  i  a \left(\frac{(2\ell-1)   Q  }{r^{2\ell+1} (Q+r^2)^3}
\frac{\left[(\ell+2) r^2 +\ell Q\right]}{4\ell(\ell+1)} +
\frac{ 1 }{4\ell Q r^{2\ell+1} }\right) Y^{(\z)}\\
\end{eqnarray}

\subsection{Second Order ($O(\epsilon^2)$)}

\begin{eqnarray}
&b& =  \frac{a^{2}}{r^{2\z}}  \left(- \frac{1}{4 r^{2} }  +
\frac{2Q+r^2 }{(Q+r^2)^2}\left( \frac{(3\z-1) - 2\z(\z+1)
\cos^2\theta}{8\z(\z+1)^2} \right)   \right) Y^{(\z)}\\
&\tilde{w}&= \frac{a^{2}}{r^{2\z} (Q+r^2)} \left(-
\frac{\z}{r^{2}} -
\frac{\cos^2\theta }{ (Q+r^2) }  \right) Y^{(\z)} \\
& b_{ra}& \equiv  b_{r}^{I_{3}}Y^{I_{3}}_{a} = \frac{a^{2}}{2r^{2\z+1}
(Q+r^2)^3 }
\left(
2\z^2 Q^2+ 3\z(\z+1) Q r^2 +  \z(\z+1) r^4\right)  \nonumber \\& &
\times\left[  \frac{\sqrt{\z}Y^{(\z+1,\z)}_{a}
}{\sqrt{(2\z+1)(\z+1)^{5} }}
+ \frac{\z-1}{2\z(\z+1)^{\frac{5}{2}}}
Y^{(\z,\z+1)}_{a} - \frac{1}{2\z^2} \sqrt{\frac{2\z-1}{\z(2\z+1) } }
Y^{(\z-1,\z)}_{a} \right]\\&b_{ua}&=b_{va}= 0
\end{eqnarray}
\begin{eqnarray}
&t_{ty}&=   -\frac{a^{2}}{4 \z
(\z+1)^2 r^{2\z} (Q+r^2)^5 }\left[ \z(\z+1)(2\z+3) Q^{3}\right.
\nonumber \\
& & +\left. \z(6\z^2+9\z+7) Q^2 r^2  + (6\z^3+ 4 \z^2+ \z + 3) Q r^4 +
(2\z^3-\z+1) r^6 \right] Y^{(l)}\nonumber \\
& & +\frac{a^{2} Q
}{2r^{2\z}(\z+1)^2(Q+r^2)^5} \sqrt{\frac{\z}{2\z+3} }\left[ (\z+1)Q^2 +
(3\z+1) Q r^2 + 2(\z+1) r^4 \right] Y^{(l)} \nonumber \\
\\
&t_{yr}&= 0,\ \ \ t_{rt}=0
\end{eqnarray}
\subsection{Third Order ($O(\epsilon^3)$)}

\begin{eqnarray}
&b &= \tilde{w} =b_{ra}=0\\
& b_{ua}& = \left(\frac{i a^{3}Q}{2(Q+r^{2})^{3}r^{2\z}(\z+1)(2\z
+3)}\sqrt{\frac{3\z(2\z+1)}{\z+2}}\right)Y^{(\z+2,\z+1)}_{a}\nonumber \\
& & +\left(\frac{i a^{3}Q}{2(Q+r^{2})^{2}r^{2\z}(\z
+1)^{\frac{3}{2}}}\left[\frac{(\z -1)(2Q+r^{2})}{4Q^2(\z +1)} - 
\frac{2\z
}{(2\z+3)(Q+r^{2})}\right] \right)Y^{(\z,\z+1)}_{a}\nonumber \\
& & -\frac{i
a^{3}}{4Q(Q+r^{2})^{3}r^{2\z+2}}\sqrt{\frac{\z}{(\z+1)^{5}(2\z+1)}}\left
[(4\z^{2}+2\z
+4)Q^{2}r^{2}+(6\z^{2}-3\z)Qr^{4}+\right. \nonumber \\
& &  \left.(2\z^{2}-\z)r^{6}
+2\z(Q+r^{2})\left((\z+1)^{2}Q^{2}-2\z Q
r^{2}-\z r^{4}\right)\right]Y^{(\z+1,\z)}_{a}\nonumber \\
& & +\frac{ia^{3}}{Qr^{2\z+2}}\sqrt{\frac{2\z
-1}{\z(2\z+1)}}\left(-\frac{1}{4(Q+r^{2})^{2}}\left((\z+1)(Q^{2}
+4Qr^{2} +2 r^{4})\right) + \frac{r^{2}}{4 l Q}\right. \nonumber \\
& & \left.+\frac{r^{2}}{8\z(\z+1)(Q+r^{2})^{3}}\left(2(2\z^{2}+3\z
-1)Q^{2} +(3 Q  r^{2} +  r^{4})(2\z^{2} +\z
-1)\right)\right)Y^{(\z-1,\z)}_{a} \nonumber \\
\end{eqnarray}
\begin{eqnarray}
& b_{va}&= -\left(\frac{i
Qa^{3}}{2(Q+r^{2})^{3}r^{2\z}}\sqrt{\frac{\z(2\z
+1)}{\z+1)}}\frac{1}{(2\z+3)(\z+1)}\right)Y^{(\z+1,\z)}_{a} \nonumber \\
& & -\left(\frac{iQa^{3}}{8\sqrt{(\z+1)^{3}}(Q+r^{2})^{3}r^{2\z
+2}}\left[2\z(\z+1)Q + \left((\z+1) +(2\z^{2}+ \z+3) \right)
r^{2}\right]\right)Y^{(\z,\z+1)}_{a} \nonumber \\
& &
+\left(\frac{iQa^{3}}{2(Q+r^{2})^{3}r^{2\z}}\frac{1}{(2\z+3)}\sqrt{\frac
{4\z}{(\z+1)(\z+2)}}\right)Y^{(\z+1,\z+2)}_{a}\nonumber \\
\end{eqnarray}
\begin{eqnarray}
     t_{yr} & = & -\left(\frac{i a^{3}Q(2\z -1)\left(\z Q +
(\z+3)r^{2}\right)}{r^{2\z
+1}(Q+r^{2})^{4}\z(\z+1)^{2}}+\frac{ia^{3}}{4r^{2\z
+3}Q(Q+r^{2})^{3}\z}\times \right. \nonumber \\
& &\left.\left[(\z +1)(Q+ r^{2})^{3} + \z(2\z -1)Q^{3} + \frac{\z(2\z
-1 )(\z +3) Q^{2}r^{2}}{\z+1}\right]\right)Y^{(\z)}\nonumber \\
& & + \frac{i a^{3}Q\left(\z Q + (\z+3)r^{2}\right)}{2r^{2\z
+1}(Q+r^{2})^{4}}\left(\frac{(2\z
-1)}{(\z+1)^{2}(\z+2)}\sqrt{\frac{\z}{(2\z +3)}}\
\right)Y^{(\z+1)}\end{eqnarray}
\begin{eqnarray}
 t_{rt} & = & \left(-\frac{ia^{3}Q\left(\z Q +
(\z+3)r^{2}\right)}{r^{2\z +1}(Q+r^{2})^{4}\z(\z+1)^{2}}+
\frac{ia^{3}}{4r^{2\z +3}Q(Q+r^{2})^{3}}\times \right.
\nonumber \\
& &\left.\left[ 1- \frac{\left((\z^{2} -1)Q^{3} +(\z^{2}-3)Q^{2}r^{2}
-3(\z +1)Qr^{4} - (\z+1)r^{6}\right)}{l(l+1)
(Q+r^2)^{3}}\right]\right)Y^{(\z)}\nonumber \\
& &+ \left(\frac{i a^{3}Q\left(\z Q + (\z+3)r^{2}\right)}{2r^{2\z
+1}(Q+r^{2})^{4}}\frac{1}{(\z+1)^{2}(\z+2)}\sqrt{\frac{\z}{(2\z+3)}}\
\right)Y^{(\z+1)} \\
t_{ty}&=&0
\end{eqnarray}

\end{appendix}

\end{document}